\documentclass[a4paper,fleqn]{cas-sc}

\usepackage[authoryear,longnamesfirst]{natbib}
\usepackage{subcaption}
\usepackage{amsthm}
\usepackage{makecell}
\usepackage[ruled,vlined]{algorithm2e}
\usepackage{float}
\newtheorem{prop}{Proposition}
\newtheorem{corollary}{Corollary}   
\def\tsc#1{\csdef{#1}{\textsc{\lowercase{#1}}\xspace}}
\tsc{WGM}
\tsc{QE}
\tsc{EP}
\tsc{PMS}
\tsc{BEC}
\tsc{DE}
\begin{document}
\let\WriteBookmarks\relax
\def\floatpagepagefraction{1}
\def\textpagefraction{.001}

% Short title
\shorttitle{Bayesian spatio-temporal weighted regression}

% Short author
\shortauthors{Junglee et~al.}

% Main title of the paper
\title [mode = title]{Bayesian spatio-temporal weighted regression for integrating missing and misaligned environmental data}                      
% Title footnote mark
% eg: \tnotemark[1]
%\tnotemark[1,2]

% Title footnote 1.
% eg: \tnotetext[1]{Title footnote text}
% \tnotetext[<tnote number>]{<tnote text>} 
%\tnotetext[1]{This document is the results of the research
%   project funded by the National Science Foundation.}

%\tnotetext[2]{The second title footnote which is a longer text matter
  % to fill through the whole text width and overflow into
  % another line in the footnotes area of the first page.}

% First author
%
% Options: Use if required
% eg: \author[1,3]{Author Name}[type=editor,
%       style=chinese,
%       auid=000,
%       bioid=1,
%       prefix=Sir,
%       orcid=0000-0000-0000-0000,
%       facebook=<facebook id>,
%       twitter=<twitter id>,
%       linkedin=<linkedin id>,
%       gplus=<gplus id>]
\author[1]{Yovna Junglee}

% Corresponding author indication
\cormark[1]

% Footnote of the first author
%\fnmark[1]

% Email id of the first author
\ead{yovna.junglee@mail.utoronto.ca}

% URL of the first author
%\ead[url]{www.cvr.cc, cvr@sayahna.org}

%  Credit authorship
\credit{Methodology, Formal Analysis, Software, Validation, Visualization, Writing – original draft}

% Address/affiliation
\affiliation[1]{organization={Department of Statistical Sciences, University of Toronto},
    %addressline={Radarweg 29}, 
    city={Toronto},
    % citysep={}, % Uncomment if no comma needed between city and postcode
    %postcode={1043 NX}, 
    % state={},
    country={Canada}}

% Second author
\author[1,2]{Vianey {Leos-Barajas}}
\credit{Methodology, Supervision, Writing – review \& editing}
% Third author
\author[1,2]{Meredith Franklin}
\credit{Conceptualization,  Funding acquisition, Investigation, Methodology, Supervision, Writing – review \& editing}
%\fnmark[2]
%\ead{cvr3@sayahna.org}
%\ead[URL]{www.sayahna.org}

%\credit{Data curation, Writing - Original draft %preparation}

% Address/affiliation
\affiliation[2]{organization={School of Environmental Sciences, University of Toronto},
    %addressline={Radarweg 29}, 
    city={Toronto},
    % citysep={}, % Uncomment if no comma needed between city and postcode
    %postcode={1043 NX}, 
    % state={},
    country={Canada}}

% Corresponding author text
\cortext[cor1]{Corresponding author}
%\cortext[cor2]{Principal corresponding author}

% Footnote text
%\fntext[fn1]{This is the first author footnote. but is common to third
 % author as well.}
%\fntext[fn2]{Another author footnote, this is a very long footnote and
%  it should be a really long footnote. But this footnote is not yet
 % sufficiently long enough to make two lines of footnote text.}

% For a title note without a number/mark
%\nonumnote{This note has no numbers. In this work we demonstrate $a_b$
%  the formation Y\_1 of a new type of polariton on the interface
%  between a cuprous oxide slab and a polystyrene micro-sphere placed
%  on the slab.
 % }

% Here goes the abstract
\begin{abstract}
Estimating environmental exposures from multi-source data is central to public health research and policy. Integrating data from satellite products and ground monitors are increasingly used to produce exposure surfaces. However, spatio-temporal misalignment often induced from missing data introduces substantial uncertainty and reduces predictive accuracy. We propose a Bayesian weighted predictor regression framework that models spatio-temporal relationships when predictors are observed on irregular supports or have substantial missing data, and are not concurrent with the outcome. The key feature of our model is a spatio-temporal kernel that aggregates the predictor over local space-time neighborhoods, built directly into the likelihood, eliminating any separate gap-filling or forced data alignment stage. We introduce a numerical approximation using a Voronoi-based spatial quadrature combined with irregular temporal increments for estimation under data missingness and misalignment. We showed that misspecification of the spatial and temporal lags induced bias in the mean and parameter estimates, indicating the need for principled parameter selection. Simulation studies confirmed these findings, where careful tuning was critical to control bias and achieve accurate prediction, while the proposed quadrature performed well under severe missingness. As an illustrative application, we estimated fine particulate matter (PM$_{2.5}$) in northern California using satellite-derived aerosol optical depth (AOD) and wildfire smoke plume indicators. Relative to a traditional collocated linear model, our approach improved out-of-sample predictive performance, reduced uncertainty, and yielded robust temporal predictions and spatial surface estimation. Our framework is extensible to additional spatio-temporally varying covariates and other kernel families.
\end{abstract}

% Use if graphical abstract is present
% \begin{graphicalabstract}
% \includegraphics{figs/grabs.pdf}
% \end{graphicalabstract}

% Research highlights
%\begin{highlights}
%\item Proposed a Bayesian weighted predictor regression framework for spatio-temporal environmental exposure modeling.
%\item Handles multi-source data with irregular supports, extensive missing values, and non-concurrent predictors.
%\item Implements Voronoi-based spatial quadrature with irregular temporal increments for fast and robust estimation under data misalignment and missingness.
%\item Compared to baseline models, shows improved PM$_{2.5}$ predictions with reduced uncertainty.
%\end{highlights}

% Keywords
% Each keyword is seperated by \sep

\maketitle

\section{Introduction}

Environmental exposures derived from multi-source data play a central role in public health. Accurate exposure estimates are needed to quantify population-level risks, guide interventions, and inform regulatory action \citep{clark2025}. Data streams such as satellite observations, ground monitors, and physical model outputs provide complementary information, but they often differ in spatial resolution, temporal frequency, and coverage \citep{lolli2025}. Integrating these sources is essential for many applications but can be methodologically challenging \citep{forlani2020}. Missing observations, spatial or temporal misalignment, and nonstationary associations between predictors and outcomes introduce uncertainty, limiting predictive reliability and the robustness of scientific inference.

Air quality research provides a clear example of these challenges. Exposure to fine particulate matter (PM$_{2.5}$) is a leading risk factor for respiratory and cardiovascular disease and premature mortality worldwide, contributing to an estimated  8.1 million deaths globally in 2021 \citep{health_effects_institute_state_2024}. While ground-based monitoring networks provide reliable measurements, they are spatially sparse and unevenly distributed \citep{zhang_satellite_2021}, limiting their ability to capture fine-scale spatial variability. 
%Accurately characterizing PM$_{2.5}$ concentrations at high spatial and temporal resolution is critical for understanding exposure patterns, guiding public health interventions, and informing environmental policy.  
Satellite-derived aerosol optical depth (AOD), available from instruments like the Moderate resolution Imaging Spectroradiometer (MODIS), offer global spatial coverage and have been widely used as proxies for surface PM$_{2.5}$ \citep{diao2019}. However, AOD retrievals are often missing due to cloud cover, snow (bright surfaces), or algorithmic failures, and when available, they are not always co-located or spatially aligned with ground monitors, leading to incomplete or misaligned data. 

To address these limitations, statistical approaches have been developed to model the PM$_{2.5}$--AOD relationship and integrate additional sources of data such as meteorology and land use characteristics. Early methods relied on linear regression \citep{wang_intercomparison_2003, chu_global_2003} and were later extended through multiple linear regression and generalized additive models to incorporate meteorological covariates \citep{van_donkelaar_estimating_2006, liu_estimating_2009, franklin2017}. In these studies, spatial misalignment was commonly addressed by interpolating or averaging AOD onto regular grids first, then modeling the PM-AOD relationship. Machine learning and deep learning have come to the forefront in interpolation efforts as well, but explicit spatio-temporal trends are not generally incorporated in these approaches \citep{li2020}. Multi-stage spatio-temporal models have been proposed to estimate daily PM$_{2.5}$ using neighbouring observations where AOD is missing, \citep{pu_spatio-temporal_2020}; however, these approaches remain limited in their ability to capture nonstationarity and locally varying relationships.  

%However, these traditional statistical approaches still struggle to capture nonstationary relationships that vary across both space and time, and cannot fully leverage local variations in environmental processes. 
Local regression techniques provide one pathway to address this gap. Geographically weighted regression (GWR) \citep{brunsdon_geographically_1998} and its spatio-temporal extension, GTWR \citep{fotheringham_geographical_2015}  apply location- and time-specific weights, allowing regression coefficients to vary across both space and time. GTWR has been shown to improve predictive performance for PM$_{2.5}$ mapping by capturing nonstationary AOD–PM$_{2.5}$ relationships \citep{he_satellite-based_2018, hu_estimating_2013}, producing fine-resolution exposure surfaces that reveal gradients and hotspots. Yet, these models still rely on concurrent predictor–outcome pairs, making them vulnerable to the systematic missingness and misalignment common in environmental datasets.
%By applying local weighting in spatio-temporal dimensions, GTWR effectively models nonstationary relationships between PM$_{2.5}$ and its predictors.  

%\cite{he_satellite-based_2018} applied GTWR to PM$_{2.5}$ mapping in China, using MODIS Dark Target and Deep Blue AOD products into a unified 3 km dataset. This approach outperformed daily GWR and two-stage models, enabling estimation for days without paired PM$_{2.5}$--AOD observations and producing high-resolution daily maps that revealed spatial gradients and hotspots. 
%Similarly, \cite{hu_estimating_2013} demonstrated that GTWR outperforms global models for PM$_{2.5}$ exposure estimation in the Southeastern U.S. Together, these studies illustrate the evolution from spatial regression models to spatio-temporal frameworks that integrate satellite data for fine-scale air quality assessment, while highlighting persistent challenges in handling missing or irregularly spaced observations.  

Other methodological advances in spatial modeling have aimed to extend beyond strictly concurrent data.  Kernel-averaged predictor frameworks \citep{heaton_spatial_2011, heaton_kernel_2012} smoothed predictors over a spatial or spatio-temporal buffer, allowing nearby information to inform estimates even when direct matches are unavailable. While promising, these  methods can be sensitive to the choice of kernel size and lag parameters, and their robustness to missing data remains unclear. This is a pressing concern in many applications, where monitoring networks are sparse, satellite retrievals are often intermittent, and episodic events such as wildfire smoke plumes create rapidly evolving exposure patterns.  

In this paper, we build on the developments of \cite{heaton_kernel_2012} by proposing a new method of estimation for the weighted predictor regression framework, tailored to settings with missing or misaligned data. Our Bayesian formulation allows flexible spatial and temporal lag structures, enabling regression even when predictors are not observed concurrently with the outcome. The framework accommodates varying buffer sizes and temporal lag specifications, and we evaluate its robustness to model misspecification as well as its performance under substantial missingness in the predictors. We applied the approach to estimate PM$_{2.5}$ in Northern California from satellite AOD and wildfire smoke plumes, quantifying local associations and demonstrating gains in predictive performance relative to traditional concurrent data models. The paper is organized as follows: Section 2 describes the methodology, provides theoretical bounds for bias induced by model misspecification, and outlines model fitting procedures. Section 3 presents a simulation study evaluating model sensitivity and robustness. Section 4 applies the methods to study spatio-temporal dependencies between daily PM$_{2.5}$ and AOD during a year where there was strong influence of wildfires on air quality in the region.

\section{Method}
\label{sec:headings}

We denote a spatio-temporal process by $\{Y(\mathbf s, t): \mathbf s \in \mathbb R^d, t \in \mathbb R\}$ where $Y(\mathbf s, t)$ represents an observation at location $\mathbf s$ and time $t$ and assume that $Y(\mathbf s, t)$ depends on another spatio-temporal temporal process $\{X(\mathbf u, \tau): \mathbf u \in \mathbb R^d, \tau \in \mathbb R\}$. In a general setting, the proposed model, similar to \cite{heaton_kernel_2012}, is given by

\begin{equation}
  \mathrm Y(\mathbf s, t) = \beta_0(\mathbf s,t) + \beta_1(\mathbf s,t)\mathrm W(\mathbf s,t) + \varepsilon(\mathbf s, t).
\end{equation}

where the weighted predictor, $\mathrm W(\mathbf s,t)$, is  
  $$\mathrm W(\mathbf s,t)=\int_{0}^q \int_{\mathcal B_r(\mathbf s)} \kappa(\mathbf s- \mathbf u,t-l\;| \;\xi(\mathbf s, t))\mathrm X(\mathbf u,t-l)d\mathbf udl$$
and $\mathcal B_r(\mathbf s) = \{\mathbf u \in \mathbb R^d \; ; \;||\mathbf s- \mathbf u||\leq r\}$. The normalization 
$\int_0^q\int_{\mathcal B_r(\mathbf s)}\kappa(\mathbf s- \mathbf u,t-l \;| \;\mathbf 
    \xi(\mathbf s, t)) d\mathbf udl =1$ ensures identifiability of $\beta(\mathbf s, t).$ In this model, each predictor $X(\mathbf{u},t-l)$  within a spatial distance $r$ of $\mathbf{s}$ and time lag $l\in[0,q]$ of $(\mathbf{s},t)$ is weighted by a spatio-temporal monotonically decreasing kernel $\kappa(\cdot\mid\xi(\mathbf{s},t))$, assumed non-negative and typically decreasing in $|\mathbf{s}-\mathbf{u}|$ and $l$. The kernel parameters $\xi(\mathbf s, t)$ may vary over space and time to allow for non-stationary weighting. Outside the region of integration, we set $\kappa(\cdot)=0$. % using some functional form or covariates.
    The noise process $\varepsilon(\mathbf s, t)$ is taken to be a zero-mean Gaussian process. The regression coefficient $\beta_1(\mathbf s,t)$ represents the effect of the weighted predictor and may vary over both space and time. 
    
This framework captures spatio-temporal variation by accounting for the local weighting of nearby predictors and recent time points, as well as the regression effect. These two sets of parameters play different roles: the (kernel) decay parameters $\xi(\mathbf s, t)$ determine the spatial extent over which the predictors can influence the parameters, whereas the regression parameters $\beta_1(\mathbf s,t)$ govern the strength of the association between the weighted predictor and the response. We model both sources of non-stationarity using smooth functions or covariates observed at the corresponding locations and times:

$$\beta_1(\mathbf s,t) = \sum b_{k}f_k(s,t) \textrm { and } \mathbb E(\xi(\mathbf s,t)) = \sum_{g=1}^G e_{g}v_g(s,t), \; \xi(\mathbf s,t) \sim \mathcal{F}(\cdot)$$

where $f_k$ and $v_g$ represent smooth basis functions (e.g. splines), or spatio-temporal covariates. When positivity is required for the kernel parameters, we  model $\log(\xi(\mathbf s,t))$ so that $\xi(\mathbf s,t) \geq 0$. 

\subsection{Choice of radius and time lag}

The choice of $r$ and $q$ must be treated as a model selection problem in which values that maximize goodness-of-fit measures such as R$^2$ and/or minimize Akaike Information Criteria (AIC) and mean-squared error (MSE) are selected. \cite{heaton_spatial_2011} choose $r$ based on the effective spatial range of the predictor. However, the effect of misspecifying  $r$ and $q$ on parameter estimation and predictive performance remains unexplored.  In this work, we focus on kernel functions of the form
\begin{equation}
    \kappa(\mathbf s- \mathbf u,t-l \; | \;\mathbf 
    \xi(\mathbf s, t)) = \kappa_1(||\mathbf s- \mathbf u ||\;| \;\mathbf 
    \xi_1(\mathbf s, t))\kappa_2(|t-l| \; | \;\mathbf 
    \xi_2(\mathbf s, t)),
    \label{eq:kappa}
\end{equation}
where $\kappa_1(\cdot)$ and $\kappa_2(\cdot)$ are  monotonically decreasing in spatial and temporal lag, respectively, and are normalized so that $\int_{\mathcal B(r)}\kappa_1(||\mathbf s- \mathbf u || \;| \;\mathbf 
    \xi_1(\mathbf s, t))\mu(d\mathbf s) =\int_{0}^q\kappa_2(|t-l| \; | \;\mathbf 
    \xi_2(\mathbf s, t))\mu(d l) =1 $.  %The former assumption holds in many practical applications whereby the effect of the covariate is expected to decrease with increasing distance.
This form is separable in the spatial and temporal lags, but because $\boldsymbol{\xi}(\mathbf s,t)$ may vary over $(\mathbf s,t)$, the overall weighting is non-stationary and not globally separable across $(\mathbf s,t)$. A fully separable kernel function arises only if $ \mathbf \xi_1(\mathbf s, t) =\mathbf \xi_1(\mathbf s)$ and $ \xi_2(\mathbf s, t) =  \xi_2(t)$, or in the simplest case if $ \mathbf \xi$ does not depend on space or time. In the following proposition, we show that the estimate of the mean of the process is biased if $r$ and $q$ are misspecified.

\begin{prop}
Let $X(\mathbf u,\tau)\ge 0$ be a spatio–temporal process and let
\[
W(\mathbf s,t; q,r,\boldsymbol\xi)
  = \int_{0}^{q} \int_{\mathcal B_r(\mathbf s)}
      \kappa_1(||\mathbf s-\mathbf u || \mid \xi_1)\,
      \kappa_2(|t-\ell| \mid \xi_2)\,
      X(\mathbf u,t-\ell)\,
      d\mathbf u\, d\ell ,
\]
where $\mathcal B_r(\mathbf s)=\{\mathbf u:\|\mathbf s-\mathbf u\|\le r\}$ and
$\kappa_1,\kappa_2$ are nonnegative kernels.
Define the true mean
\[
\mu(\mathbf s,t)=\beta_0 +\beta_1\, W(\mathbf s,t; q,r,\boldsymbol\xi)
\]
and the misspecified mean
\[
\tilde\mu(\mathbf s,t)=\beta_0+\beta_1\, W(\mathbf s,t; \tilde q,\tilde r,\boldsymbol\xi).
\]

Then the difference between the two mean functions satisfies
\[
\begin{aligned}
|\mu(\mathbf s,t)-\tilde\mu(\mathbf s,t)|
 &\le
 \beta_1 \int_{0}^{\max\{q,\tilde q\}}
     \int_{\mathcal A_{r,\tilde r}(\mathbf s)}
         \kappa_1(||\mathbf s-\mathbf u||)\,
         \kappa_2(|t-\ell|)\,
         X(\mathbf u,t-\ell)
     \,d\mathbf u\, d\ell \\[0.8em]
 &\quad +
 \beta_1 \int_{\mathcal B_{\max\{r,\tilde r\}}(\mathbf s)}
        \int_{\min\{q,\tilde q\}}^{\max\{q,\tilde q\}}
            \kappa_1(||\mathbf s-\mathbf u||)\,
            \kappa_2(|t-\ell|)\,
            X(\mathbf u,t-\ell)
        \,d\ell\, d\mathbf u ,
\end{aligned}
\]
where  
\[
\mathcal A_{r,\tilde r}(\mathbf s)
  = \{\mathbf u : \min(r,\tilde r)\le \|\mathbf s-\mathbf u\|\le \max(r,\tilde r)\}
\]
is the spatial annulus between the two radii.
\end{prop}

Proposition 1 implies that the discrepancy between the true and misspecified means is influenced not only by the behaviour of the kernel functions but also by the magnitude of the process outside the true spatial–temporal window. The proof of this proposition can be found under Appendix C of the Supplementary Material. 

%\begin{prop}
%    same prior from an exponential family for both. The posterior estimates $E_Y(E(\tilde \beta | \xi,\cdot)) \to E_Y(E( \beta | \xi, \cdot))$ if $\delta_1\to 0$ and $\delta_2 \to 0.$
%\end{prop}

%\begin{proof}
%\end{proof}

\begin{corollary}
Suppose there exist constants $\delta_1, \delta_2 > 0$ such that
\[
\kappa_1(||\mathbf s-\mathbf u||) \le \delta_1 \quad \text{for all } \|\mathbf s-\mathbf u\| > r,
\qquad
\kappa_2(|t-\ell|) \le \delta_2 \quad \text{for all } |t-\ell| > q.
\]

Then, for any $\tilde r \ge r$ and $\tilde q \ge q$, the difference between the true and misspecified means satisfies
\begin{align*}
\bigl|\mu(\mathbf s,t)-\tilde\mu(\mathbf s,t)\bigr|
&\le
\beta_1 \left(\delta_1\delta_2
\int_0^{\tilde q} \int_{\mathcal A_{r,\tilde r}(\mathbf s)} 
    X(\mathbf u,t-\ell)\, d\mathbf u\, d\ell
\;+\;
\delta_1\delta_2 \int_q^{\tilde q} \int_{\mathcal B_{\tilde r}(\mathbf s)} 
    X(\mathbf u,t-\ell)\, d\mathbf u\, d\ell\right) \\[0.5em]
&\le
\beta_1\bigg(\delta_1\delta_2\,\tilde q\, \pi (\tilde r^2 - r^2)\, \sup_{(\mathbf u,\ell)\in \mathcal A_{r,\tilde r}(\mathbf s)\times [0,q]} X(\mathbf u,t-\ell)
\;+\;\\ &
\delta_1\delta_2\, (\tilde q - q)\, \pi \tilde r^2 \, \sup_{(\mathbf u,\ell)\in \mathcal B_{\tilde r}(\mathbf s)\times [q,\tilde q]} X(\mathbf u,t-\ell)\bigg).
\end{align*}
\end{corollary}

From collorary 1, if $\delta_1,\delta_2 \to 0$, then the bound only depends on the behaviour of $X(\mathbf u, t-l)$ i.e.\ as long as $X(\mathbf u, t-l)$ does not contain extremes, the error bound $\to 0$. %That is, choosing a large enough radius and temporal lag could reduce bias. 
In this setting, one might consider selecting $\tilde{r}$ and $\tilde{q}$ such that $\mathcal{B}(\tilde{r})\times[0,\tilde q] = \mathcal{D} \times \mathcal{T}$, thereby covering the entire spatio-temporal domain under consideration. While this choice appears reasonable in theory, it is often computationally prohibitive in practice. For example, on a regular 1 km² daily grid, the number of data points required to cover a region of radius $\mathcal{B}(\tilde{r})\times[0,\tilde q]$ is $\pi \tilde{r}^2 \tilde{q}$, which grows quadratically with $\tilde{r}$. Thus, it is recommended that model selection be based on out-of-sample performance assessed under a set of varying spatial and temporal lag values.

\subsection{Model estimation}

\subsubsection{Numerical approximation of spatio-temporal integral}
In practice, we only observe the spatio-temporal processes at discrete time-points and locations: $\{Y(s_n, t_i) \; :i \in \{1,\ldots,\mathrm{T_{Y_n}}\}, n \in \{1,\ldots,\mathrm{N_Y}\}\}$, and  $\{X(s_m, t_j) \; :j \in \{1,\ldots,\mathrm{T_{X_m}}\}, m \in \{1,\ldots,\mathrm{N_X}\}\}$.

%Denote the vector 
Define $\tilde{\mathbf X}_{ni} = \{X(\mathbf s_m,t_j) \; : \; ||\mathbf s_m- \mathbf s_n||\leq r, \; t_i-t_j\leq q \}$ and $\mathbf z_{ni} = \{\mathbf s_m \;:\;||\mathbf s_m- \mathbf s_n||\leq r \}$ and $\mathbf d_{ni} = \{( \mathbf s_m,t_j)\;:\;||\mathbf s_m-\mathbf s_n||\leq r, \; t_i-t_j\leq q \}$ with $|\tilde{\mathbf X}_{ni} | = |\mathbf d_{ni}|=M_{ni}<\infty$, then, the objective is to approximate $\mathrm W(\mathbf s,t)$ from $\mathrm W(\mathbf s_n,t_i)$ and fit

\begin{equation}
            Y(\mathbf s_n, t_i) = \beta_0(\mathbf s_n, t_i) + \beta_1(\mathbf s_n, t_i)\mathrm W(\mathbf s_n, t_i\;; \;\mathbf 
    \xi(\mathbf s_n, t_i)) + \varepsilon(\mathbf s_n, t_i).
\end{equation}

When data are missing or misaligned, this estimation is  more challenging. \cite{heaton_spatial_2011} use Monte Carlo integration and \cite{heaton_kernel_2012}  project $X(\mathbf s, t)$ onto a regular grid and approximate the integral by summing pointwise products of the discretised covariate with its weights after fitting a latent Gaussian process - an approach that can be computationally expensive for finely spaced data due to covariance estimation. We instead use a numerical approximation for the weighted predictor function:;

$$\mathrm W(\mathbf s_n,t_i) = \sum_{k}^{M_{ni}}\left(\kappa(d_{ni,k} \;| \;\mathbf 
    \xi(\mathbf s_n, t_i))\cdot\tilde{\mathbf X}_{ni,k}\right) \cdot\Delta\tilde{\mathbf X}_{ni,k}$$

where $\Delta\tilde{\mathbf X}_{ni,k} = \textrm{area}(V_{ni,k})\cdot \delta_k$. Denote the $k^{th}$ observation in the buffer by $\tilde{\mathbf X}_{ni,k} = X(\mathbf s_k,t_k)$, then $V_{ni,k}$ is the corresponding area of the cell from the Voronoi tessellation of the spatial location $\mathbf s_k$. Suppose $X(s_k,\cdot)$ is observed at time-points $(t_{1},\ldots,t_{k-1},t_k,\ldots,T_{X_k})$ then $\delta_k=t_k-t_{k-1}$.

A Voronoi tessellation is a computational technique that partitions a metric space based on proximity to a given set of points \cite{okabe_spatial_2009}. At time $i$, given a finite set of distinct points given by $\mathbf z_{ni}$ which represents the locations at which we have observed data within the buffer of location $n$, the Voronoi cell $V_{ni,k}$ associated with location $z_{ni,k}$ is defined as:
$$
V_{ni,k} = \left\{ x \in \mathcal B_r(\mathbf s_n)\mid \|x - z_{ni,k}\| \leq \|x - z_{ni,g}\|, \ \forall g \neq k \right\}.
$$

Each cell $V_{ni,k}$ consists of all points in the plane that are closer to $z_{ni,k}$ than to any other site.  The union of all such cells forms a partition of the buffer region. The area of each Voronoi cell $V_{ni,k}$ quantifies the spatial influence or weight of the corresponding observation point, based on proximity to neighbouring points.

Because $Y$ and $X$ are typically collected on different space-time supports, the choice of $r$ and $q$ directly affects $M_{ni}$, the number of covariate values contributing to $W(\mathbf s_n,t_i)$. In applications such as PM$_{2.5}$ estimation from satellite AOD, there can be substantial missing data in $X(\mathbf v, \tau)$ that yield $M_{ni} =0$ (or very small), in which case $Y(\mathbf s_n,t_i)$ is poorly approximated. In this case, $Y(\mathbf s_n,t_i)$ should be removed to avoid large approximation error. In Appendix A, we illustrate the Voronoi tessellation of the buffer on two different days. The tessellation depends on the location of the observed AOD data. An advantage of this method is that it accounts for regions where AOD is unobserved by approximating them using the nearest observed values, rather than treating them as missing, as is often assumed in traditional collocation approaches.
%We can lower the number of such unmatched observations by increasing $r$ and/or $q$ however this comes with increasing computational costs to compute the weighted covariate in particular when the spatio-temporal predictor process is observed on a fine spatial grid.

%\subsubsection{Description of Markov chain Monte Carlo algorithm}
\subsubsection{Bayesian inference}

A Metropolis-within-Gibbs sampler is used to perform inference. We assume a Gaussian for prior $b_k \sim \mathcal N(0,\sigma^2_b)$. This choice of prior is conditionally conjugate under a Gaussian process assumption for $\varepsilon(\mathbf s,t) \sim \mathcal{GP}(0,\sigma^2 C(\mathbf s,\mathbf s^\prime,t ,t^\prime))$. A conditionally conjugate prior  $\mathcal{IG}(a_b, b_b)$ is assumed for $\sigma_b^2$. Similarly, a conditionally conjugate non-informative inverse-Gamma prior is assumed for $\sigma^2 \sim \mathcal{IG}(a, b)$.  The prior for $\mathbf \xi(\mathbf s,t)$ depends on the choice of covariance function and assumed model. In general for $\mathbf \xi(\mathbf s,t) = \xi$ , we propose an informative Gamma $\sim \mathcal{G}(c_1,c_2)$ for non-spatio-temporal scale parameters governing the kernel functions particularly in the presence of missing data. A natural choice for $c_1$ and $c_2$ is based on the expected rate of decay over which the predictor is allowed to extend. Under this prior specification, a Metropolis-Hasting step is used to sample from the posterior distribution. Alternative choice of priors for different parameterizations of the scale parameters is discussed further in Section 4. Full details of the Metropolis-within-Gibbs sampler implementation are available in Appendix D.

\section{Simulation Study}
We conducted a simulation study to understand the behaviour of the model under different conditions. First we generated the spatio-temporal process via:

$$\log(X_t(\mathbf s)) = \mu_t(\mathbf s) + \nu_t(\mathbf s)$$
with
\begin{eqnarray}
\mu_t(\mathbf{s}) = & \exp\left( -\left( \frac{(s_1 - s_{1,1,t})^2}{2 \cdot 50^2} + \frac{(s_2 - s_{2,1,t})^2}{2 \cdot 10^2} \right) \right) 
+ 1.5  \exp\left( -\left( \frac{(s_1 - s_{1,2,t})^2}{2 \cdot 70^2} + \frac{(s_2 - s_{2,2,t})^2}{2 \cdot 15^2} \right) \right) +  \\& 0.5  \exp\left( -\left( \frac{(s_1 - s_{1,3,t})^2}{2 \cdot 40^2} + \frac{(s_2 - s_{2,3,t})^2}{2 \cdot 8^2} \right) \right)
\end{eqnarray}

and $\nu_t(\mathbf s)$ is a zero-mean spatial Gaussian field assuming an exponential covariance model $
C(h) = \sigma^2 \exp\left(-\frac{h}{\phi}\right)$
with variance \(\sigma^2 = 1\) and scale (range) parameter \(\phi = 0.1\). Here $(s_{1,k,t},s_{2,k,t})$ denotes the centre of the $k$th (k=1,2,3) peak at time $t$; the three peak locations are redrawn at each time point. %The choice of the mean governs the position of three randomly selected peaks at each time point. 
We then generate the data using 
    $$\mathrm Y(\mathbf s, t) = \beta_0 + \beta_1\int_{0}^5 \int_{\mathcal B_{10}(\mathbf s)} c_1c_2\exp \bigg(-\frac{||\mathbf s- \mathbf u||}{2}\bigg)\exp\bigg(-\frac{|t- l|}{0.5
}\bigg)\mathrm X(\mathbf u,t-l)d\mathbf udl + \varepsilon$$
where $\varepsilon \sim \mathcal N(0, 0.25^2)$ and $c_1$ and $c_2$ are normalizing constants for the spatial and temporal kernels so that each integrates to 1 over its support. In this model specification, only predictors within a 10 km radius and the most recent 5 time lags affect the response ($r=10, \; q=5$). As shown in the illustrations of the simulated surface and predictor process provided in Appendix B, the spatio-temporal weights rapidly decay ($\to 0$) for $r>10$ and $q>3$. Finally, we set $\beta_0=10$ and $\beta_1=0.5$. 

\begin{figure}
    \centering
    \includegraphics[width=\linewidth]{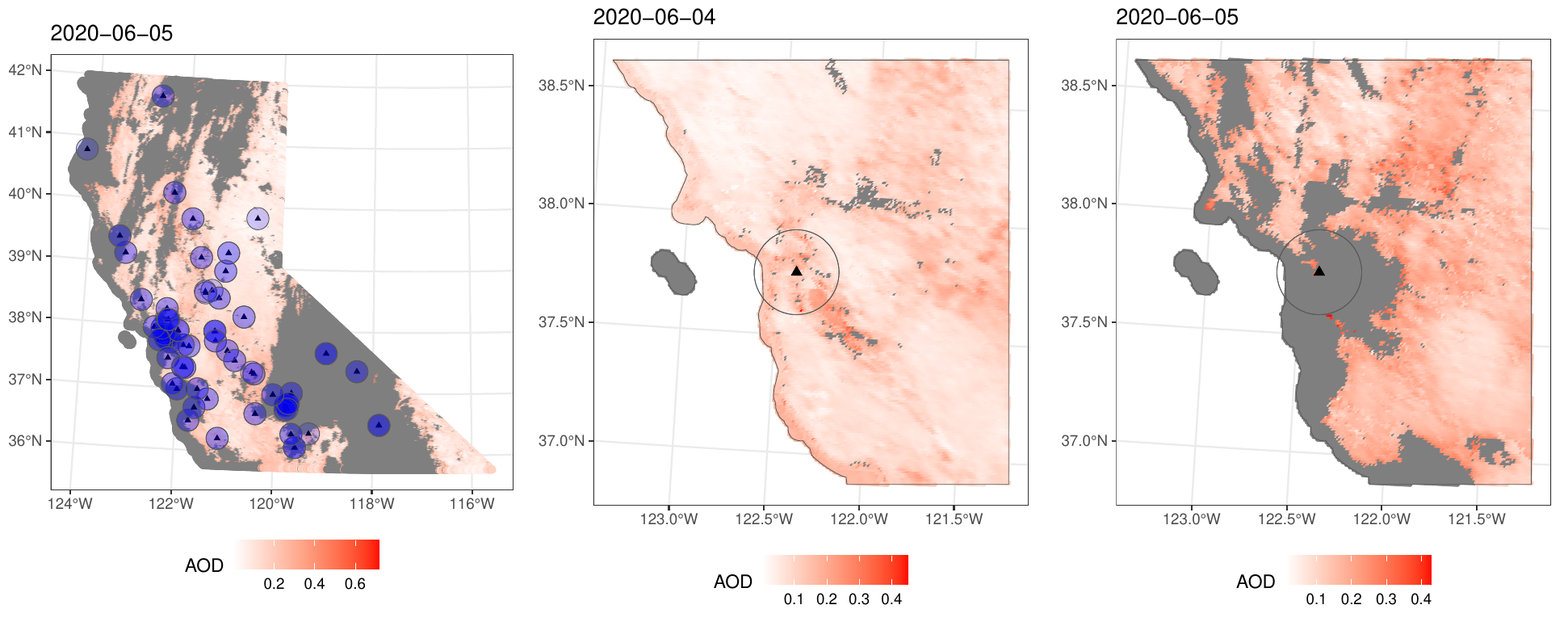}
    \caption{Map of Northern California study region on two consecutive days in June 2020 showing: (a) PM$_{2.5}$ monitoring locations (blue dots) with 20 km buffers around them (shaded blue circles) with 1 km x 1 km MAIAC aerosol optical depth (AOD) grid (pink to red); (b) zoom in of one PM$_{2.5}$ monitor on a day with nearly completely observed AOD within 20 km; and (c) zoom in of one PM$_{2.5}$ monitor on the next day with significant missing AOD within 20 km.}
    \label{fig:buffers}
\end{figure}

We generated 100 replicated data sets on N = 100 locations and T = 153 time points. We retained N = 25 locations for model fitting and the remaining out-of-sample to compare predictive performance. This choice reflects the sparseness of ground-level stations in practice. Using the fully observed predictor process, we fit (i) a model with the true lag values $(r,q)=(10,5)$, (ii) a truncated misspecified model with $\tilde r = 5 < r$ and $\tilde q = 2 < q$ (iii) an extended misspecified model with $\tilde r = 15 > r$ and $\tilde q = 8 > q$. The covariate process $X_t(\mathbf s)$ is well-behaved (no extremes). To assess the approximation under irregular sampling, we also simulated datasets with missing values that mimic the application's missingness pattern observed (described in Figure \ref{fig:buffers}). For inference, we found that using a non-informative inverse-Gamma prior,
$
\sigma^2 \sim \mathcal{IG}(a_1, b_1),
$
does not pose issues for parameter estimation or identifiability when the data are complete. However, in the presence of missing data, employing an informative prior helps improve both identifiability and sampler mixing.

\begin{figure}
    \centering
    % First subfigure
    \begin{subfigure}{\textwidth}
        \centering
        \includegraphics[width=\linewidth]{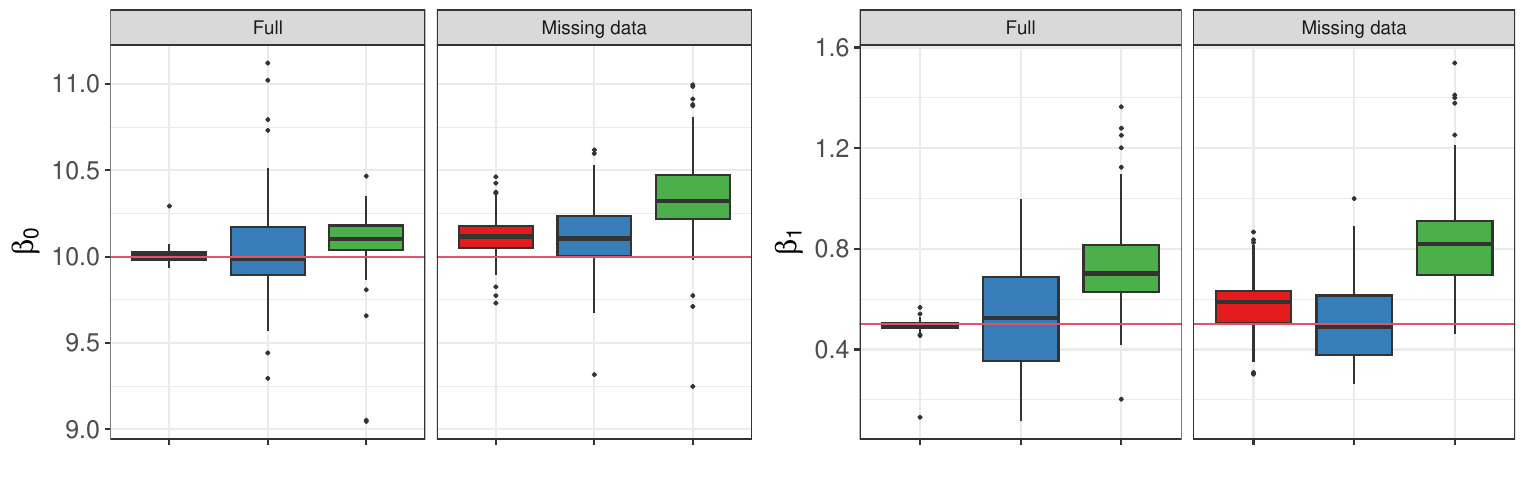}
       % \caption{First subfigure}
        \label{fig:sub1}
    \end{subfigure}
    \hfill
    % Second subfigure
    \begin{subfigure}{\textwidth}
        \centering
        \includegraphics[width=\linewidth]{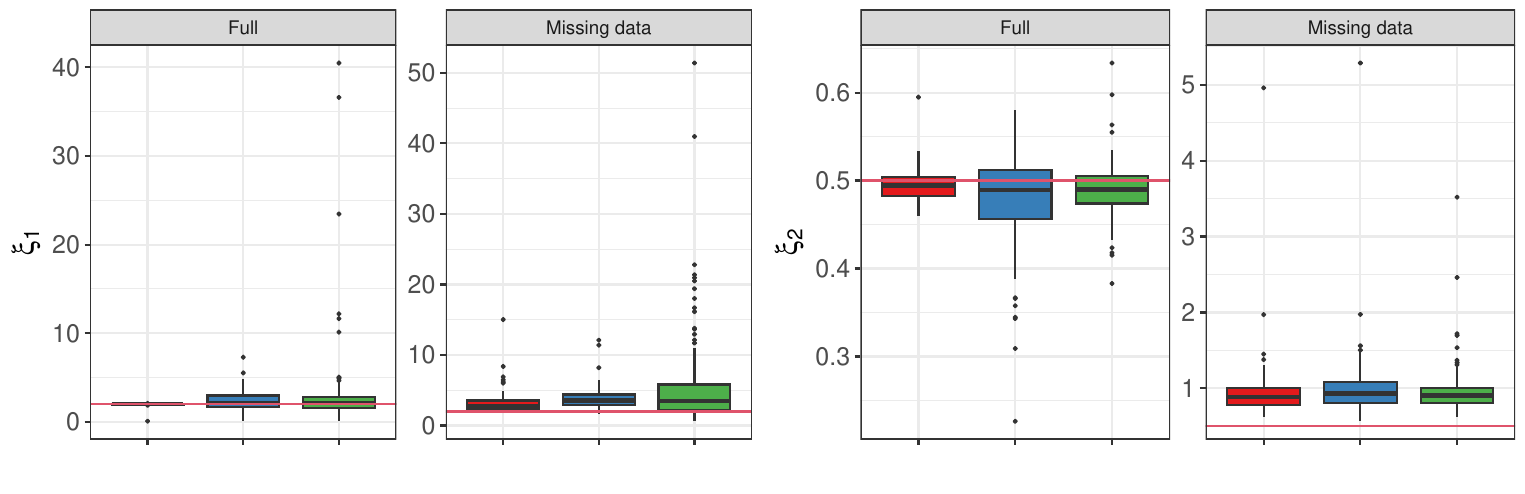}
        %\caption{Second subfigure}
        \label{fig:sub2}
    \end{subfigure}
 \caption{Distribution of estimated model parameters across 100 simulated data sets with complete data (full) and missing data. Red boxplots indicate the results obtained using true radius and time lag values $(r=10, q=5)$, while blue $(\tilde{r}=15, \tilde{q}=8)$ and green $(\tilde{r}=5, \tilde{q}=2)$ illustrate the variability of estimates under misspecified radius and time lags.}
    \label{fig:simparam}
    \end{figure}
Figure ~\ref{fig:simparam} illustrates the effect of model misspecification on parameter estimation. When $\tilde{r} < r$ and $\tilde{q} < q$, the model is more prone to overestimating parameters. In particular, we observe large biases in the estimated  scale parameter of the spatial kernel function. This pattern appeared both with full data setting and under missingness. A larger value or $\xi_1$ corresponds to a more uniform weighting across the radius (i.e. slower decay). Consequently, selecting a smaller radius can lead the model to overestimate the influence of predictors within $\tilde r$ to compensate for omitted contributions from $\mathcal A_{\tilde r, r}(\mathbf s)$. Similarly, the effect of parameters $\beta_0$ and $\beta_1$ is overestimated to offset the mean bias induced by the reduced integration domain. However, $\xi_2$ is estimated more consistently because the true data-generating process assumes a faster temporal decay than the lag parameter used to generate the data. However, because estimates of the scale parameters are sporadic, stronger priors should be favoured if model inference and interpretability are of primary concern. With missing data, the numerical approximation introduces some bias; even with the true $(r,q)$, parameter estimates are positively biased and the bias is typically exacerbated under model misspecification. 
%The top left panel in Figure \ref{fig:simoos} displays the distribution of the noise parameter $\sigma^2$. Misspecification of the mean process results in systematically higher estimates of $\sigma^2$. In the presence of missing data, estimated $\sigma^2$ is generally larger, reflecting increased uncertainty.

   \begin{figure}
     % Second subfigure
    %\begin{subfigure}{\textwidth}
     %   \centering
     %   \includegraphics[width=\linewidth]%{plots/sigR2.pdf}
       % \caption{Second subfigure}
    %    \label{fig:r2}
   % \end{subfigure}
        % Second subfigure
  %  \begin{subfigure}{\textwidth}
        \centering
        \includegraphics[width=\linewidth]{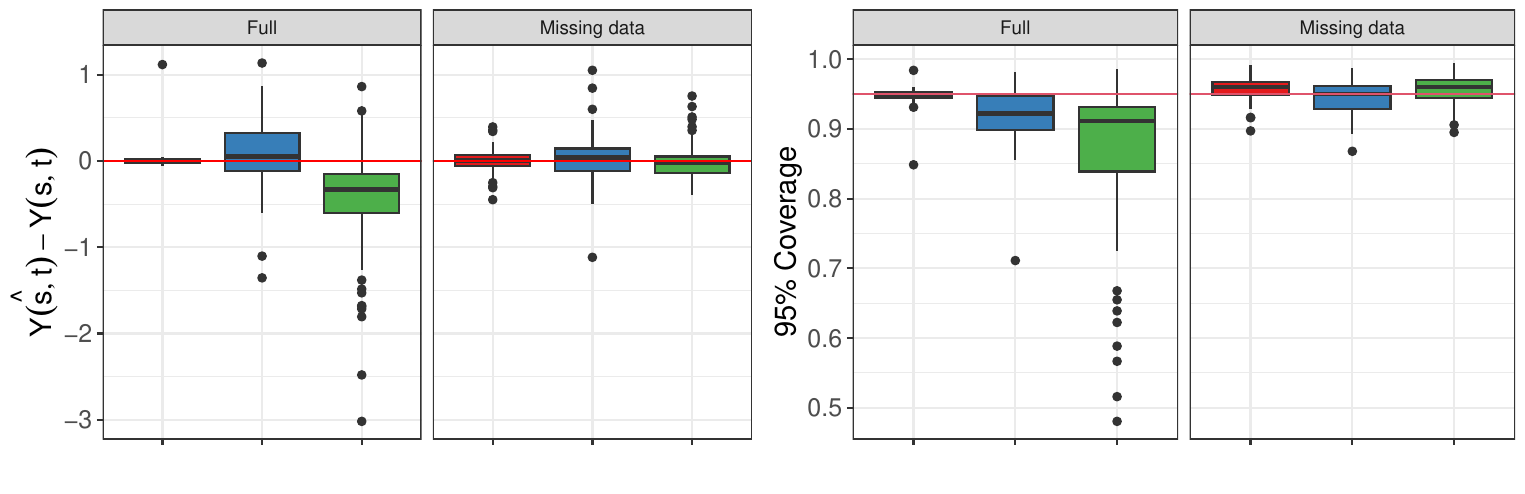}
       % \caption{Second subfigure}
        \label{fig:mse}
   % \end{subfigure}
    \caption{Out-of-sample performance of simulated data measured using the mean deviation of the predictions from the true values, and coverage of the 95\% prediction interval. Red boxplots indicate the results obtained using true radius and time lag values $(r=10, q=5)$, while blue $(\tilde{r}=15, \tilde{q}=8)$ and green $(\tilde{r}=5, \tilde{q}=2)$ illustrate the effect of model misspecification on predictive performance.}
    \label{fig:simoos}
\end{figure}

Under the full data setting, the out-of-sample deviations ($\hat Y(\mathbf s,t)-Y(\mathbf s,t)$) results in Figure \ref{fig:simoos} align with Corollary 1. When the kernel functions decay to zero within the true region of integration, selecting larger $\tilde r$ and $\tilde q$ reduces the mean bias compared to a smaller integration region. Additionally, predictions are underestimated when choosing smaller lags with lower coverage of the 95\% prediction intervals. In the presence of missing data, we observe similar patterns in the deviations of the predictions from the true value. Most importantly, our proposed numerical approximation produces robust model predictions with high coverage of the prediction intervals. These results show that in practice choosing  sufficiently large lag values ensures both reliable parameter estimation and predictions.

%This pattern persists in the presence of missing data. Similarly, the mean-squared error is low relative to the performance under the true parameters, although misspecified models exhibit greater variability in out-of-sample performance. Under the full data setting, coverage of the reduced model is lower, with prediction interval widths ranging from 2.81 to 6.30, compared to the wider intervals of the extended model (2.75 to 8.44). Finally, even with missing data, out-of-sample performance remains strong, with coverage exceeding 95\%, reflecting wider prediction intervals and higher uncertainty estimates.
 
\section{Application to PM\textsubscript{2.5} mapping in Northern California}

\subsection{Data}
The dataset comprises of fine particulate matter (PM\textsubscript{2.5}) measurements from 67 air quality monitoring stations across Northern California between 1st June 2020 and 31st October 2020, coinciding with the region’s peak wildfire season. Ground-based data were linked with satellite data of aerosol optical depth (AOD) from the Multi-Angle Implementation of Atmospheric Correction (MAIAC) algorithm, which provides high-resolution (\~ 1 km x 1 km) gridded observations from an algorithm applied to observations from the Moderate Resolution Imaging Spectroradiometer (MODIS) instrument aboard NASA’s Terra and Aqua satellites \citep{lyapustin2018}. To account for wildfire smoke influence on PM$_{2.5}$ concentrations, we included smoke plume data from the National Oceanic and Atmospheric Administration’s (NOAA) Hazard Mapping System (HMS), which delineates smoke plume extent based on satellite imagery \citep{NOAA_HMS_Smoke}. Together these data enabled the characterization of spatiotemporal variation in PM\textsubscript{2.5} and its association with wildfire smoke during the 2020 fire season in Northern California.

\subsection{Model Specifications and Estimation}

In Table 1, we present the different models fitted to the data. The spatiotemporal baseline model assumes exponential kernels for both spatial and temporal lags. For the spatial kernel scale parameter we used a Gamma prior of the form $\mathcal G(a,b)$ where $a=1, b=9$ with the shape-scale parametrization $E(\xi_1) = 9$ placing mass on spatial scales consistent with AOD influence extending to 20 km. For the temporal kernel scale we used a weakly informative inverse-Gamma prior $\xi_2\sim\text{IG}(a_0,b_0)$.

Building on the baseline model, the functional mean model augments the intercept by a temporally varying mean function $\beta_0(t)$ represented with B-spline basis functions $f_k(t)$ \citep{wood_generalized_2017}. We used $K=10$ equally spaced knots to capture bi-weekly trends, and retained the same priors as in the baseline model.

In the remaining models we included a varying coefficient for wildfire smoke. Let $I(\mathbf s ,t)$ indicate a smoke plume event (1 = satellite-derived wildfire smoke plume intersected a PM$_{2.5}$ monitor $Y(\mathbf s ,t)$ at location $s$ on day $t$, 0 = otherwise). We incorporated this indicator to the models: 
(1) at the mean level where $\beta_0(\mathbf s,t) = \sum_k b_{0k} f_k(t)+\gamma I(\mathbf s, t)$ assuming a normal prior for $\gamma$, (2) as a varying coefficient where the effect of the weighted predictor of AOD varied by plume status $\beta_1(\mathbf s,t)=\beta_{10}+\beta_{11}I(\mathbf s,t)$. We fix $\xi_2$ at the posterior median from the functional mean model and allowing the effect of the kernel-averaged AOD to depend on smoke plume status. This was motivated by \cite{zhang_inconsistent_2004}, who showed that range and scale parameters in covariance functions are not consistently estimable and require stabilization. %To facilitate model estimation, we fix $\xi_2$ to the posterior median of $\xi_2$ obtained in the functional mean model and estimate the remaining parameters under the same prior assumptions. 

The full model includes the varying coefficient specification above and allows the spatial extent over which AOD influences PM$_{2.5}$ to vary based on smoke plumes via $\log\{\xi_1(\mathbf s,t)\}=\alpha_0+\alpha_1\,I(\mathbf s,t)+\nu,\qquad \nu\sim N(0,\sigma_\nu^2)$. To help estimation, when $I(\mathbf s,t)=0$, we placed a strong point-mass prior on $\xi_1(\mathbf s,t)$ centered at the posterior median of the samples obtained under the varying coefficient model. We assume normal priors for $\mathbf \alpha$ and inverse-gamma for $\sigma_\nu^2$. Under this approach, we effectively only estimated the effect of $I(\mathbf s,t)$ through $\alpha_1$. The MCMC algorithms used for inference can be found under Appendix D in the Supplementary Material. Convergence was assessed using the Gelman-Rubin test. Each model was run for 25000-50000 MCMC iterations, with the first 50\% of the samples used as warm-up and retaining the latter half for inference.  As a benchmark, we also fit a collocated linear model where we aggregated and averaged AOD data within a 5 km radius around each station on each day. 

% Define a centered p-column type
\newcolumntype{C}[1]{>{\centering\arraybackslash}p{#1}}

\begin{table}[!b]
\caption{Spatio-temporal models for PM$_{2.5}$: mean specification, temporal and spatial regression coefficients, and scale parameters.}
\small
\renewcommand{\arraystretch}{1.2}
\begin{tabular}{
    C{3.0cm}
    C{2.25cm}
    C{2.25cm}
    C{3.75cm}
    C{3.0cm}
}
\toprule
\textbf{Model} & $\beta_0(\mathbf{s},t)$ & $\beta_1(\mathbf{s},t)$ & $\xi_1(\mathbf{s},t)$ & $\xi_2(\mathbf{s},t)$ \\
\midrule
Baseline &
$\beta_0$ &
$\beta_1$ &
$\xi_1 \sim \mathrm{Gamma}(1,9)$ &
$\xi_2 \sim \mathrm{IG}(a_0,b_0)$ \\

Functional mean  &
$\sum_k b_{0k}\, f_k(t)$ &
$\beta_1$ &
$\xi_1 \sim \mathrm{Gamma}(1,9)$ &
$\xi_2 \sim \mathrm{IG}(a_0,b_0)$ \\

Varying coefficient  &
\makecell{$\sum_k b_{0k}\, f_k(t)$ \\ $+~\gamma\, I(\mathbf s,t)$} &
\makecell{$\beta_1(\mathbf{s},t)=\beta_{10}$ \\ $+~\beta_{11} I(\mathbf s,t)$} &
$\xi_1 \sim \mathrm{Gamma}(1,9)$ &
\makecell{$\xi_2$ fixed at \\ posterior median$^\dagger$} \\

Full  &
\makecell{$\sum_k b_{0k}\, f_k(t)$ } &
\makecell{$\beta_1(\mathbf{s},t)=\beta_{10}$ \\ $+~\beta_{11} I(\mathbf s,t)$} &
\makecell{$\log\!\{\xi_1(\mathbf s,t)\}=\alpha_0$ \\ $+~\alpha_1 I(\mathbf s,t)+\nu,$ \\ $\nu\sim N(0,\sigma_\nu^2)$} &
\makecell{$\xi_2$ fixed at \\ posterior median$^\dagger$} \\

Collocated Linear &
$\beta_0$ &
$\beta_1$ &
-- & -- \\
\bottomrule
\end{tabular}

\footnotesize $^\dagger$\,Following \cite{zhang_inconsistent_2004}, $\xi_2$ is fixed at the posterior median from the functional mean model to stabilize range estimation. In all models, $\beta_1(\mathbf s,t)$ multiplies $W(\mathbf s,t)$, the kernel-weighted AOD predictor defined in Eq.(1), which averages $X(\mathbf u,t-l)$ over $\mathcal B_r(\mathbf s)\times[0,q]$ using normalized exponential kernels in space and time. The collocated linear benchmark model uses the AOD average $\overline{X}_{5\text{km}}(\mathbf s,t)$ in 5 km buffer instead.
\end{table}

We assumed independent Gaussian errors as our primary objective is to  characterize the spatio-temporal dependence of PM$_{2.5}$ on AOD. We intentionally excluded additional meteorological covariates to isolate this relationship and better understand the spatio-temporal patterns driven by AOD and wildfire smoke presence.

\subsection{Results}

The performance of each method is assessed using the root-mean square error (RMSE) metric, continuous ranked probability score (CRPS) to evaluate the accuracy of predictive forecasts \citep{gneiting2007strictly} and the coverage of the 95\% prediction interval for stations that were kept out-of-sample. 

The collocated linear model produced an RMSE of 25.8, CRPS 6.89 and coverage 0.941. To fit this  model, 407 observations were removed due to unpaired PM$_{2.5}$-AOD observations.%$R^2 = 0.184$

Out-of-sample predictive performance of the competing models is reported in Table~\ref{tab:results}. Overall, the larger radius around each site (r = 20 km) had better predictive accuracy (lower RMSE and CRPS), except for the full model, which performed better with r = 10 km. When $r = 20$ the varying coefficient model achieved the best balance across evaluation metrics, outperforming the alternative model specifications. %These findings highlight the sensitivity of model performance to the choice of buffer radius and emphasize the importance of carefully selecting this parameter in practice. %In particular, the full model performs noticeably worse when the radius is set to $r = 20$, suggesting that the added complexity at this radius may induce overfitting and reduce generalizability. However, we observed that the error is largely driven by overinflated predictions at a single location; when that location is removed, the model performance improves to RMSE = 19.3, CRPS =  6.81 and coverage = 0.942.
%R$^2$ = 0.455,.

\newcolumntype{C}[1]{>{\centering\arraybackslash}p{#1}}
\begin{table}[H]
\caption{Spatio-temporal model predictive performance for $r\in\{10,20\}$ km and $q=3$.}
\label{tab:results}

\renewcommand{\arraystretch}{1.2}
{\setlength{\tabcolsep}{4pt}
\begin{tabular}{
    C{4.0cm}
    C{1.7cm}
    C{1.7cm}
    C{1.7cm}
    C{1.7cm}
    C{1.7cm}
    C{1.7cm}
}
\toprule
& \multicolumn{3}{c}{$r=10,\;q=3$} & \multicolumn{3}{c}{$r=20,\;q=3$} \\
\cmidrule(lr){2-4}\cmidrule(lr){5-7}
\textbf{Model} & \textbf{RMSE} & $\textbf{CRPS} $ & \textbf{Coverage} & \textbf{RMSE} & $\textbf{CRPS} $ & \textbf{Coverage} \\
\midrule
Baseline & 36.3& 8.49 & {0.957} & {36.1} &  {8.22}  & 0.953  \\
Functional mean& {19.6} & 6.32 & 0.942 & 20.1 & {6.21} & {0.945} \\
\makecell{Varying coefficient} & 23.5 & 6.97 & 0.942 & \textbf{19.1} & \textbf{6.07} & \textbf{0.946} \\
Full  & {19.6} & {6.22} & 0.943 & 20.1 & 6.34 & {0.944} \\
\bottomrule
\end{tabular}}
\flushleft\footnotesize Root mean square error (RMSE) units in $\mu$g/m$^3$; collocated linear model: RMSE = 25.8 $\mu$g/m$^3$, CRPS = 6.89, Coverage = 0.941. Values in bold indicate the lowest RMSE and CRPS and the highest coverage across all models.
\end{table}

%\newcolumntype{C}[1]{>{\centering\arraybackslash}p{#1}}
%\begin{table}[H]
%\caption{Spatio-temporal model predictive performance for $r\in\{10,20\}$ km and $q=3$.}
%\label{tab:results}

%\renewcommand{\arraystretch}{1.2}
%{\setlength{\tabcolsep}{4pt}
%\begin{tabular}{
%    C{4.0cm}
%    C{1.7cm}
%    C{1.7cm}
%    C{1.7cm}
%    C{1.7cm}
%    C{1.7cm}
%    C{1.7cm}
%}
%\toprule
%& \multicolumn{3}{c}{$r=10,\;q=3$} & \multicolumn{3}{c}{$r=20,\;q=3$} \\
%\cmidrule(lr){2-4}\cmidrule(lr){5-7}
%\textbf{Model} & \textbf{RMSE} & $\boldsymbol{R^2}$ & \textbf{Coverage} & \textbf{RMSE} & $\boldsymbol{R^2}$ & \textbf{Coverage} \\
%Baseline & \textbf{32.5} & \textbf{0.230} & \textbf{0.956} & 35.1 & 0.227 & 0.957 \\
%Functional mean& \textbf{19.6} & \textbf{0.464} & \textbf{0.940} & 21.0 & 0.410 & 0.942 \\
%\makecell{Varying coefficient} & 21.0 & 0.362 & 0.944 & \textbf{19.9} & \textbf{0.400} & %\textbf{0.944} \\
%Full  & \textbf{19.0} & \textbf{0.454} & \textbf{0.940} & 63.3 & 0.037 & 0.940 \\
%\bottomrule
%\end{tabular}}
%\footnotesize Collocated baseline model: RMSE = 25.8, $R^2=0.184$, Coverage = 0.941.
%\end{table}

\begin{figure}
     % Second subfigure
    \begin{subfigure}{.31\textwidth}
        \centering
        \includegraphics[width=\linewidth]{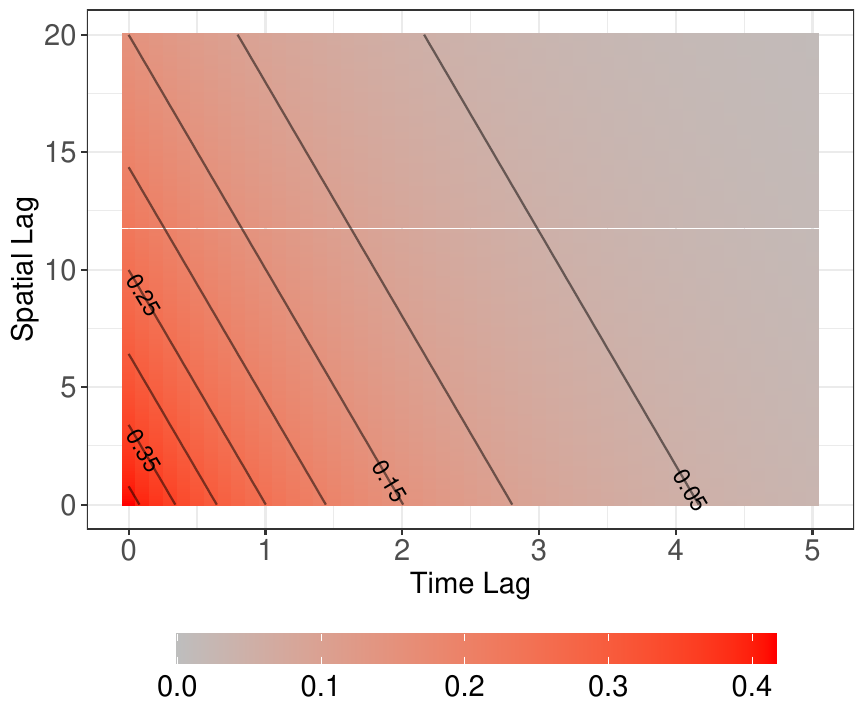}
        \caption{Functional mean \newline ($r=20$)}
        \label{fig:func20s}
    \end{subfigure}%
        %Second subfigure
    \begin{subfigure}{.31\textwidth}
        \centering
        \includegraphics[width=\linewidth]{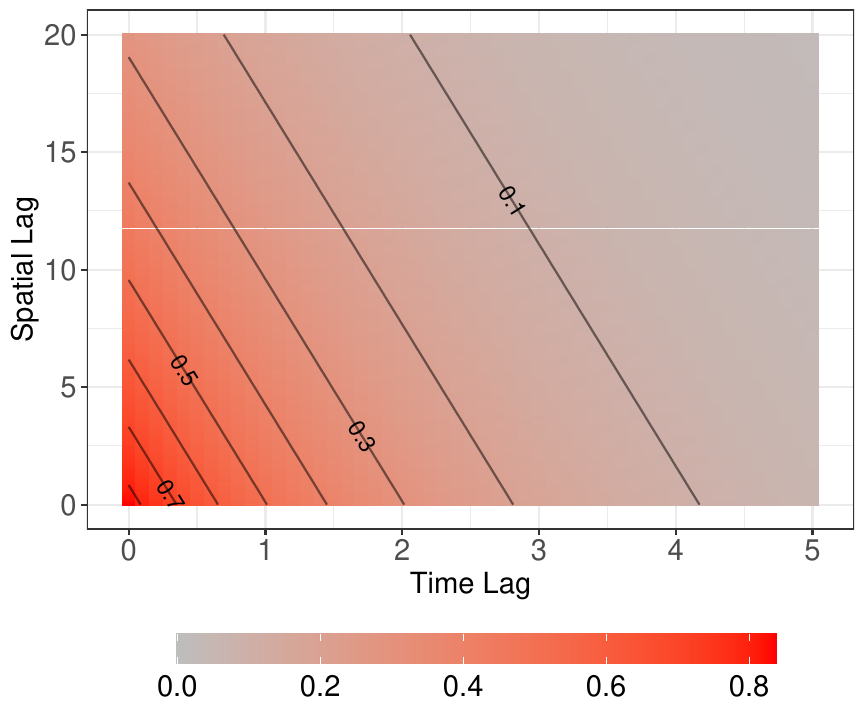}
        \caption{Varying coefficient model in the \newline absence of fire plumes ($r=20$) }
        \label{fig:vc20f0}
    \end{subfigure}%
      \begin{subfigure}{.31\textwidth}
        \centering
        \includegraphics[width=\linewidth]{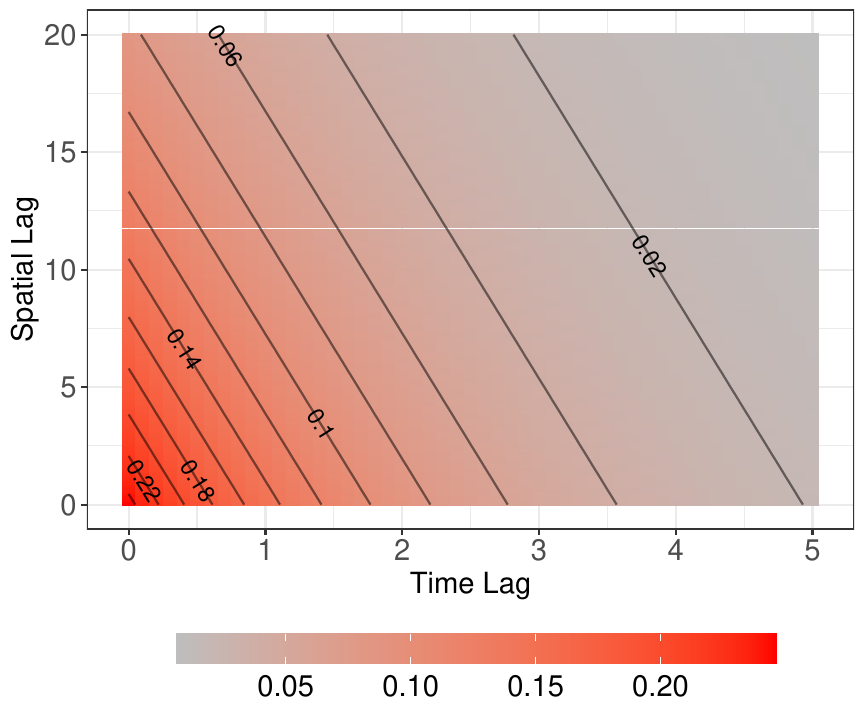}
       \caption{Varying coefficient model in the presence of fire plumes ($r=20$) }
        \label{fig:vc20f1}
    \end{subfigure}
    \centering
    \begin{subfigure}{.31\textwidth}
        \centering
        \includegraphics[width=\linewidth]{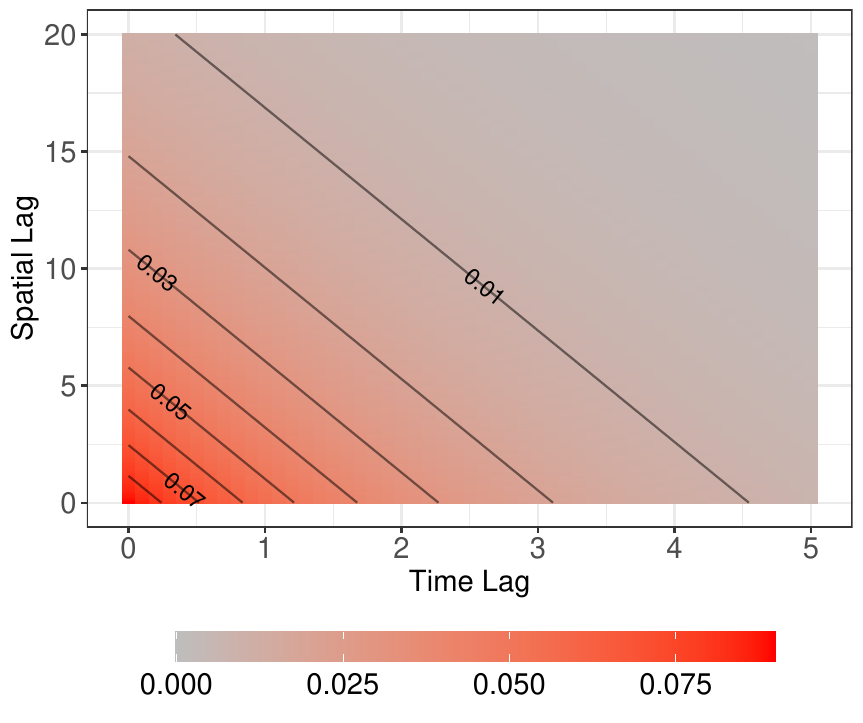}
        \caption{Full model in the absence of\newline fire plumes ($r=10$) }
        \label{fig:full10f0}
    \end{subfigure}%
      \begin{subfigure}{.31\textwidth}
        \centering
        \includegraphics[width=\linewidth]{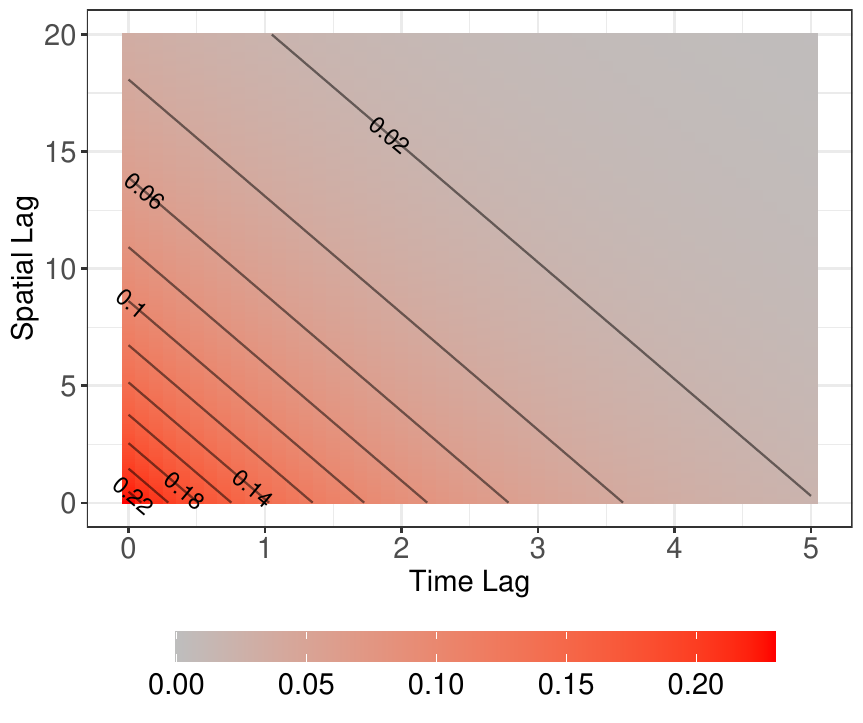}
       \caption{Full model in the presence of fire plumes ($r=10$) }
        \label{fig:full10f1}
    \end{subfigure}
    \caption{Estimates of the surface explaining the spatio-temporal dependencies between PM2.5 and AOD at varying spatial and time lags. }
    \label{fig:surfest}
\end{figure}

To gain further insight into the spatio-temporal dependencies between PM$_{2.5}$ and AOD, Figure~\ref{fig:surfest} illustrates the estimated surface 
$
\hat{\omega}(h,l) = \hat{\beta}_1 \, \kappa_1(h; \hat{\xi}_1(\mathbf{s}, t)) \, \kappa_2(l; \hat{\xi}_2(\mathbf{s}, t)),
$
for the three best-performing models: the functional mean model (Figure~\ref{fig:func20s}), the varying coefficient model (Figures~\ref{fig:vc20f0} and~\ref{fig:vc20f1}), both with $r=20$, and the full model (Figures~\ref{fig:full10f0} and~\ref{fig:full10f1}) with $r=10$. We used the posterior medians of the MCMC samples to construct the surface $\widehat{\omega}(h, l)$. Figures~\ref{fig:vc20f0} and ~\ref{fig:full10f0} show the estimated effect of AOD on PM$_{2.5}$ when no smoke plume was present ($I(\mathbf{s},t)=0$), whereas Figure~\ref{fig:vc20f1} and ~\ref{fig:full10f1} show the effect under plume conditions ($I(\mathbf{s},t)=1$). The results indicate that incorporating wildfire smoke information alters the spatio-temporal weighting. The varying coefficient model is able to isolate the effect of AOD under different plume conditions. In the presence of fire plumes, the model estimates higher levels of PM$_{2.5}$ with $\hat \gamma \in [0.461, 0.587]$ on average while the effect of the kernel-averaged AOD being smaller and more localised in its immediate neighbourhood. The full model captures this differently, the magnitude of the effect of the weighted AOD on PM$_{2.5}$ is higher in the presence of smoke plumes. These results suggest that fire events amplify the local influence of AOD on PM$_{2.5}$, underscoring the importance of accounting for such events when constructing accurate exposure maps.

%in the functional mean model, the AOD effect was distributed more evenly across the buffer region, while in the varying coefficient model, both for $I(\mathbf{s},t)=0$ and $I(\mathbf{s},t)=1$, the effect of AOD decayed more rapidly over space and time but attained larger magnitudes in the immediate neighbourhood of an observation.   

  \begin{figure}
     % Second subfigure
    \begin{subfigure}{\textwidth}
        \centering
        \includegraphics[width=\linewidth]{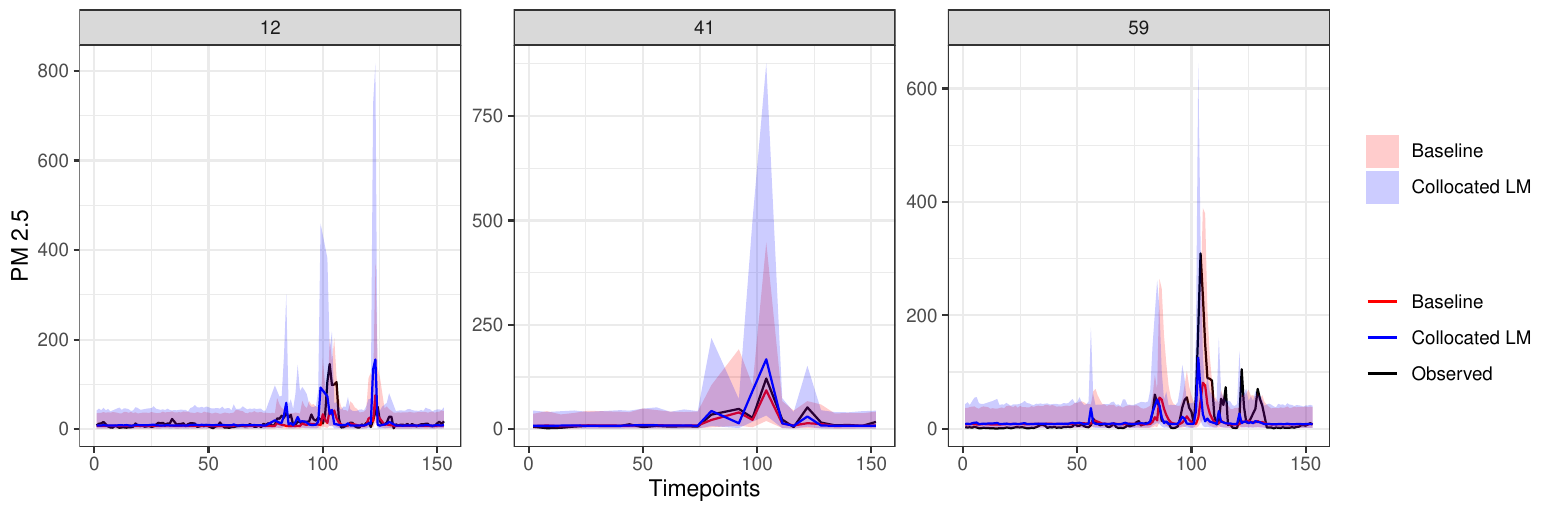}
        \caption{}
        \label{fig:pred1}
    \end{subfigure}
        %Second subfigure
    \begin{subfigure}{\textwidth}
        \centering
        \includegraphics[width=\linewidth]{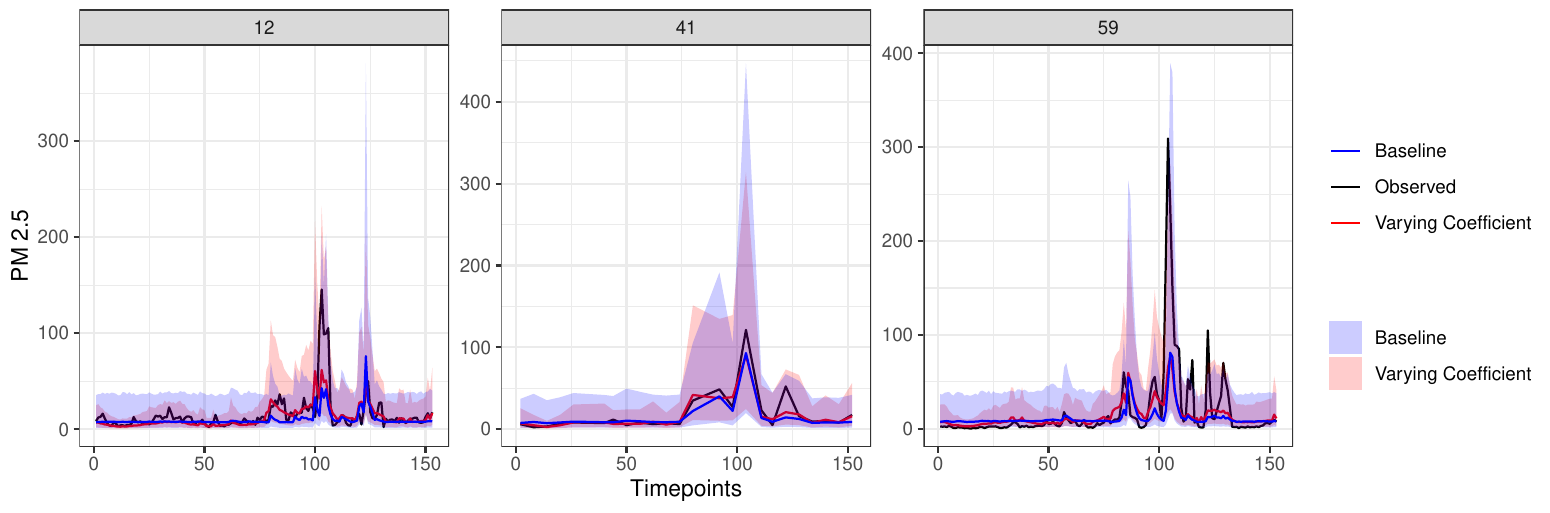}
        \caption{}
        \label{fig:pred2}
    \end{subfigure}
    \caption{Predictions of PM$_{2.5}$ at out-of-sample locations over time. The top panel compares the collocated linear model with the baseline model ($r=20$). The bottom panel compares the baseline model with the varying coefficient model ($r=20$). As model complexity increases, uncertainty quantification of the predictions improves with narrower prediction intervals capturing peaks.}
    \label{fig:pred}
\end{figure}

Figure~\ref{fig:pred} compares out-of-sample temporal predictions and 95\% posterior prediction intervals at held-out monitoring stations. The collocated linear benchmark model attained reasonable point predictions but produced very wide intervals, especially during sharp concentration peaks. By contrast, the baseline model has substantially narrower interval widths while maintaining coverage, providing more precise uncertainty quantification for PM$_{2.5}$ concentrations across both peak episodes and stable periods. Figure~\ref{fig:pred2} further shows that incorporating wildfire information improves temporal calibration. The full model produced narrower prediction intervals that still captured the observed peaks, highlighting the importance for wildfire influences in PM$_{2.5}$ estimation.  

\begin{figure}
    \centering
    \includegraphics[width=\linewidth]{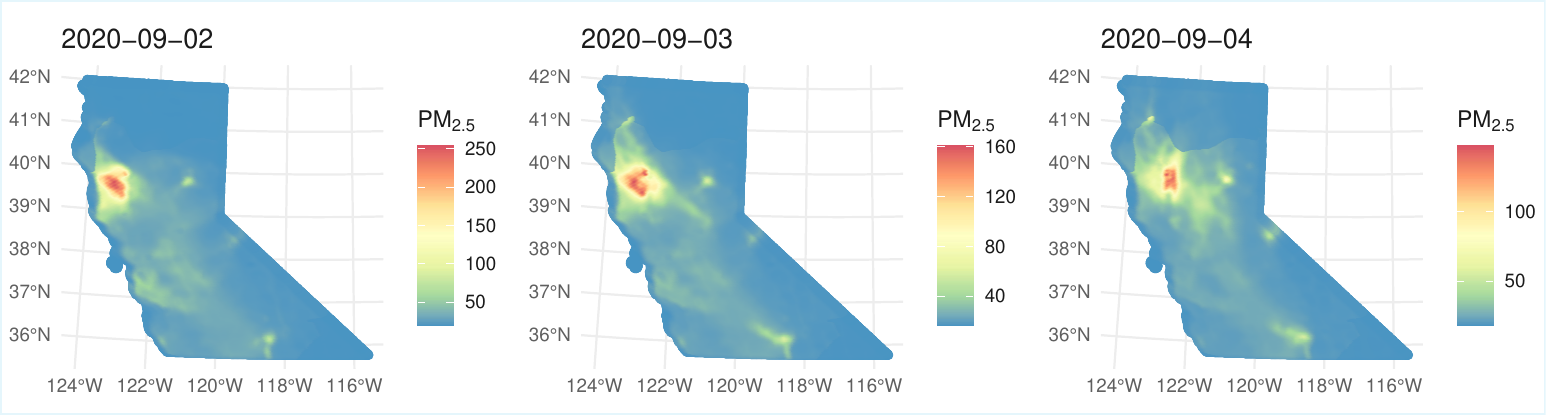}
    \caption{Spatial mapping of PM$_{2.5}$ using the full model ($r=10$) for three consecutive days in September 2020.}
    \label{fig:mapspred}
\end{figure}

Figure~\ref{fig:mapspred} presents 1 km resolution gridded maps of estimated PM$_{2.5}$ derived from 1~km resolution AOD for three days in September during peak wildfire season, using the functional mean model. The approach effectively captured both spatial gradients and day-to-day temporal patterns, producing smooth, coherent variation across space and time, particularly where the August Complex wildfire occurred. An advantage of our approach is it enables fine-resolution predictions even when AOD is missing.

\section{Discussion}

Handling spatial misalignment traditionally relies on change-of-support or hierarchical modeling to reconcile point- and areal-level data \citep{gotway_young_2002}. In point-to-point and point-to-grid misalignment, common practice is to average data from the misaligned source over buffer regions around a target point to align (i.e. forced collocation). This approach can clearly induce misclassification for several reasons including choice of the buffer radius. Furthermore, misalignment may be present due to missing data. Our contribution is to embed the missing data problem inside a normalized spatio-temporal weighting operator whose kernel parameters can vary with covariates, coupled with a numerically stable quadrature that respects missingness and irregular supports. This yielded a unified route to estimation and uncertainty quantification without a separate imputation stage, while remaining computationally tractable.

The weighted kernel-averaged predictor regression model we developed required careful specification of the spatial and temporal lag windows $(r,q)$ for valid inference and accurate prediction. Our theoretical results quantify how misspecification of $(r,q)$ propagated to bias in the mean, which in turn lead to biased parameter estimates and poorer predictive performance. Our proposed quadrature scheme that uses Voronoi-based spatial weights and irregular time steps to estimate the spatio-temporal integral, enabled estimation under extensive missing data and spatio-temporal misalignment. Unlike common strategies that first gap-fill (interpolate) and then regress on the complete interpolated surface (e.g. \cite{li2014}), our approach builds the weighting and uncertainty handling directly into the likelihood via $W(\mathbf s,t)$, allowing propagation of uncertainty from the available data to the predictions.

Missingness and and spatial misalignment are common issues in many environmental modeling applications. In the satellite AOD–PM$_{2.5}$ literature, missing AOD due to clouds, snow, high aerosol loadings, or retrieval failures is widespread, and many studies use explicit gap filling via interpolation or ensembles before regression \citep{xiao2021, sorek-hamer2020,li2020}. When we applied our approach to estimate PM$_{2.5}$ from satellite AOD over Northern California in 2020, which had an intense wildfire season, the weighted predictor model captured both spatial and temporal variation and yielded narrower, better-calibrated prediction intervals than a collocated linear baseline. Incorporating smoke plume information altered the estimated weighting surface and improved temporal calibration around concentration peaks. The proposed estimator handled missing and misaligned data effectively because it aggregated over whatever nearby $(\mathbf u,l)$ were observed, rather than requiring a fully gridded covariate. When no informative neighbours existed within $(r,q)$, the observation is flagged rather than imputed, avoiding potentially biased reconstructions.

Some limitations of our work remain. First, we focused on the PM$_{2.5}$-AOD relationship and intentionally excluded meteorology to isolate this signal and develop methodology that targets the spatial misalignment and missing data issue. However, in practice models often adjust for other predictors to account for residual confounding from meteorological variables such as boundary-layer height, temperature, humidity, and winds. Thus, a natural extension to our work is to include spatio-temporally varying meteorology either as additional weighted predictors (with their own kernels and possibly interactions) or as modulators of the AOD kernel parameters $\boldsymbol\xi(\mathbf s,t)$ (e.g., humidity-dependent temporal decay or wind-aligned spatial kernels). Second, we used exponential kernels; alternative families (e.g., Matern-like or compactly supported kernels) and nonseparable forms could better match plume transport physics. Third, for stability we fixed the temporal range in some specifications; fully hierarchical estimation of all kernel scales with stronger priors or penalization is an important avenue. Fourth, we assumed conditionally independent Gaussian errors; other residual structures (e.g., nonseparable errors) may further improve calibration, however may unnecessarily increase model complexity at the cost of computational intensity. Finally, scaling our models to larger spatial domains could computationally benefit from sparse quadrature, multi-resolution bases, or approximate inference (e.g. via INLA) tailored to kernel-weighted predictors.

Overall, the weighted predictor framework provides a robust and flexible tool for spatio-temporal exposure modeling beyond strictly concurrent effects, with particular strength under missing and misaligned covariates. Future work will integrate multi-source covariates (meteorology, emissions, chemical transport model outputs) and explore adaptive, plume-aware kernels to further enhance predictive performance and interpretability.

\printcredits

%% Loading bibliography style file
% \bibliographystyle{model1-num-names}
\bibliographystyle{cas-model2-names}

% Loading bibliography database
\bibliography{cas-refs}

\newpage

\appendix

\section{Voronoi Tessellation}

\begin{figure}
     % Second subfigure
    %\begin{subfigure}{\textwidth}
     %   \centering
     %   \includegraphics[width=.33\linewidth]{plots/vor1.pdf}
   % \end{subfigure}
        % Second subfigure
   % \begin{subfigure}{.45\textwidth}
     %   \centering
     %   \includegraphics[width=\linewidth]{plots/vor2.pdf}
   % \end{subfigure}
    %\begin{subfigure}{.45\textwidth}
        \centering
        \includegraphics[width=.5\linewidth]{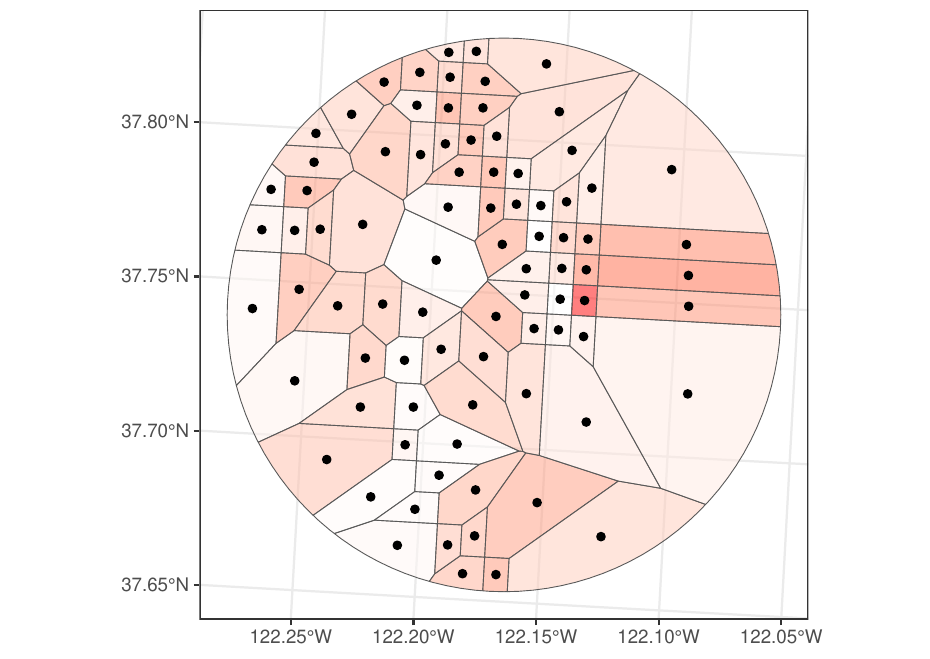}
   % \end{subfigure}
    \caption{Example of a Voronoi tessellation over a buffer.}
    \label{fig:voreg}
\end{figure}

Figure ~\ref{fig:voreg} shows the Voronoi tessellation over a buffer of 20km for different days. On days where there is missing AOD data, the region is weighted more heavily due to a larger Voronoi cell.

\section{Simulation Study}

%Figures \ref{fig:xfield} and \ref{fig:surfsim} display the true surface and simulated random fields for the spatio-temporal predictor process. 

\newpage

 %   \begin{figure}
     % Second subfigure
    %\begin{subfigure}{\textwidth}
     %   \centering
     %   \includegraphics[width=.33\linewidth]{plots/vor1.pdf}
   % \end{subfigure}
        % Second subfigure
%    \begin{subfigure}{.45\textwidth}
%        \centering
%        \includegraphics[width=\linewidth]{plots/x1.pdf}
%    \end{subfigure}
%    \begin{subfigure}{.45\textwidth}
%        \centering
 %       \includegraphics[width=\linewidth]{plots/x2.pdf}
 %  \end{subfigure}
%    \caption{Examples of simulated $X(\mathbf u,l)$ %fields.}
 %   \label{fig:xfield}
%\end{figure}
    \begin{figure}
        \centering
        \includegraphics[width=.5\linewidth]{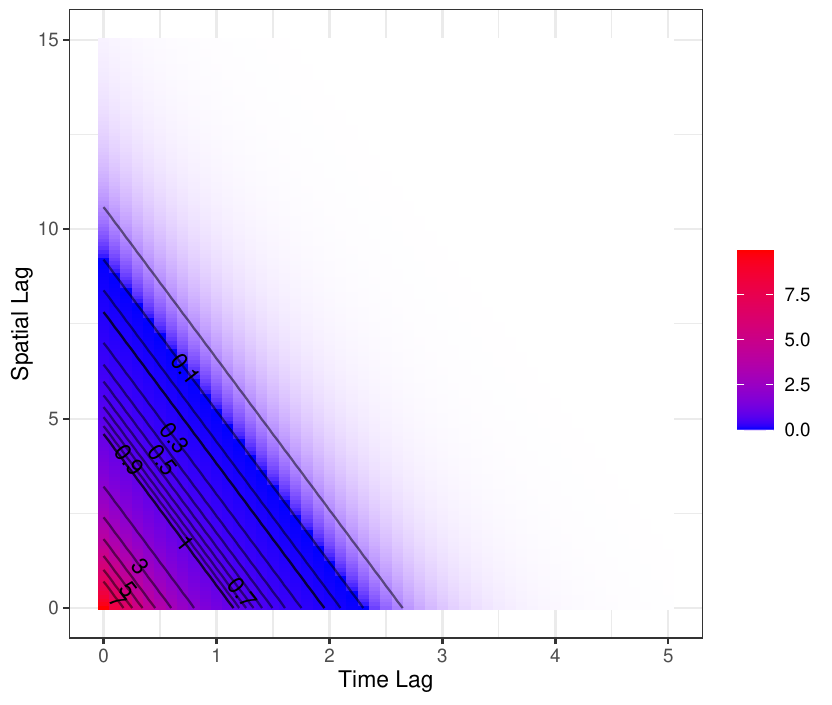}
    \caption{Examples of weight surface used in simulation study.}
    \label{fig:surfsim}
\end{figure}

\newpage

\section{Proof of Proposition 1}
\begin{proof}
We consider the four possible cases.

\begin{enumerate}

\item[\textnormal{(1)}] $\tilde r \ne r$ and $\tilde q = q$.  
The two spatial balls differ, but the time window is the same.  
The symmetric–difference region is the annulus  
\[
\mathcal A_{r,\tilde r}(\mathbf s)
  = \{\mathbf u : \min(r,\tilde r)\le \|\mathbf s-\mathbf u\|\le \max(r,\tilde r)\}.
\]
Thus
\[
\bigl|\mu(\mathbf s,t)-\tilde\mu(\mathbf s,t)\bigr|
  \le \beta_1 \int_{0}^{q} \int_{\mathcal A_{r,\tilde r}(\mathbf s)}
      \kappa_1(\mathbf s-\mathbf u)\,\kappa_2(t-\ell)\,
      X(\mathbf u,t-\ell)\, d\mathbf u\, d\ell .
\]

\item[\textnormal{(2)}] $\tilde q \ne q$ and $\tilde r = r$.  
The spatial region is the same, but the time windows differ.  
The symmetric–difference region is the extra time slab  
\[
[\min(q,\tilde q),\,\max(q,\tilde q)]
\]
over the common ball $\mathcal B_r(\mathbf s)$.  
Hence
\[
\bigl|\mu(\mathbf s,t)-\tilde\mu(\mathbf s,t)\bigr|
  \le \beta_1 \int_{\mathcal B_r(\mathbf s)}
      \int_{\min(q,\tilde q)}^{\max(q,\tilde q)}
      \kappa_1(\mathbf s-\mathbf u)\,\kappa_2(t-\ell)\,
      X(\mathbf u,t-\ell)\, d\ell\, d\mathbf u .
\]

\item[\textnormal{(3)}] $\tilde r > r$ and $\tilde q > q$.  
The region included by the larger window consists of  
\[
A = \bigl(\mathcal A_{r,\tilde r}(\mathbf s)\times[0,q]\bigr)
    \;\cup\;
    \bigl(\mathcal B_{\tilde r}(\mathbf s)\times[q,\tilde q]\bigr).
\]
The bounding region is
\[
B = \bigl(\mathcal A_{r,\tilde r}(\mathbf s)\times[0,\tilde q]\bigr)
    \;\cup\;
    \bigl(\mathcal B_{\tilde r}(\mathbf s)\times[q,\tilde q]\bigr).
\]
Since $A\subset B$ and $\kappa_1,\kappa_2,X$ are nonnegative,
the integral over $A$ is bounded by the integral over $B$, giving the desired inequality.

\item[\textnormal{(4)}] $\tilde r < r$ and $\tilde q < q$.  
The argument is analogous.  
The omitted region is
\[
A = \bigl(\mathcal A_{\tilde r,r}(\mathbf s)\times[0,\tilde q]\bigr)
    \;\cup\;
    \bigl(\mathcal B_r(\mathbf s)\times[\tilde q,q]\bigr),
\]
and the bounding region
\[
B = \bigl(\mathcal A_{\tilde r,r}(\mathbf s)\times[0,q]\bigr)
    \;\cup\;
    \bigl(\mathcal B_r(\mathbf s)\times[\tilde q,q]\bigr)
\]
satisfies $A\subset B$.  
Nonnegativity of the integrand again yields the bound.

\item[\textnormal{(5)}] $\tilde r > r$, $\tilde q < q$.  
The mismatch arises from a \emph{larger} spatial radius but a \emph{smaller} temporal lag.

\smallskip
(i) The part omitted by the shorter temporal window is
\[
A_1 = \mathcal B_r(\mathbf s)\times[\tilde q,\,q].
\]

(ii) The part added by enlarging the spatial radius is the annular region
\[
A_2 = \mathcal A_{r,\tilde r}(\mathbf s)\times[0,\,\tilde q].
\]

Thus the absolute difference of the two integrals equals the integral of the
(nonnegative) integrand over $A_2$ minus the integral over $A_1$ in magnitude, and
is therefore bounded above by the integral over
\[
A_1 \cup A_2.
\]

The given bounding region is
\[
B
  = \bigl(\mathcal A_{r,\tilde r}(\mathbf s)\times[0,\,q]\bigr)
    \;\cup\;
    \bigl(\mathcal B_{\tilde r}(\mathbf s)\times[\tilde q,\,q]\bigr).
\]

Since \(A_1\cup A_2 \subset B\) and the integrand
\(\kappa_1\,\kappa_2\,X\) is nonnegative, the difference
\(|\mu(\mathbf s,t)-\tilde\mu(\mathbf s,t)|\) is bounded by the integral over \(B\).

\item[\textnormal{(6)}] $\tilde r < r$, $\tilde q > q$.  
The mismatch arises from a \emph{larger} temporal lag but a \emph{smaller} spatial radius.

\smallskip
(i) The part omitted by the smaller radius is
\[
A_1 = \mathcal A_{\tilde r, r}(\mathbf s)\times[0,\,q].
\]

(ii) The part added by increasing temporal lag
\[
A_2 = \mathcal B_{\tilde r}(\mathbf s)\times[q,\tilde q].
\]

Thus the absolute difference of the two integrals equals the integral of the
(nonnegative) integrand over $A_1$ minus the integral over $A_2$ in magnitude, and
is therefore bounded above by the integral of the integrand over
\[
A_1 \cup A_2.
\]

The given bounding region is
\[
B
  = \bigl(\mathcal A_{\tilde r,r}(\mathbf s)\times[0,\tilde q]\bigr)
    \;\cup\;
    \bigl(\mathcal B_{r}(\mathbf s)\times[q,\tilde q]\bigr).
\]

Since \(A_1\cup A_2 \subset B\) and the integrand
\(\kappa_1\,\kappa_2\,X\) is nonnegative, the difference
\(|\mu(\mathbf s,t)-\tilde\mu(\mathbf s,t)|\) is bounded by the integral over \(B\).
\end{enumerate}
\end{proof}

\newpage
\section{Brief description of MCMC algorithm} \label{sec:MCMC_Appendix}
Let $\beta$ denote the regression coefficients $[ \mathbf b, \gamma]$ Under a Gaussian likelihood, the conditional posterior distribution $\beta|\cdot$ is given by
\begin{align*}
\mathbf \beta|\mathbf \xi,\sigma^2 \propto \pi(\mathbf\beta)\times\prod_{i,j}f(Y(s_i,t_j)\; ;W(s_i,t_j \;; \xi(s_i,t_j)),
\sigma^2) \tag{A1.1}
\end{align*}
where $f(\cdot \; ;\mu, \sigma^2)$ is the p.d.f of a Gaussian distribution. Under a Gaussian prior distribution $\mathcal N(0,\sigma_b^2)$, $\mathbf \beta|\cdot$ is conditionally normally distributed. We assume a conditionally conjugate inverse-gamma prior for $\sigma_b^2$,
\begin{align*}
    \sigma_b^2 | \cdot \propto \pi(\sigma_b)\times f(\beta\;; 0,\sigma_b^2)\tag{A1.2}
\end{align*}

Similarly, the conditional posterior distribution $\sigma^2|\cdot$ is given by,
\begin{align*}
\mathbf \sigma^2|\mathbf \xi,\mathbf \beta \propto \pi(\sigma^2)\times\prod_{i,j}f(Y(s_i,t_j)\; ;W(s_i,t_j \;; \xi(s_i,t_j)), \tag{A2}
\sigma^2). 
\end{align*}
We assign the prior distribution $\sigma^2 \sim \mathcal{IG}(0.001,0.001)$ which results in a conditional posterior distribution for $\sigma^2|\cdot$ that also follows an inverse-gamma distribution.

1. For the baseline, functional mean and varying coefficient models where $\xi_1(\mathbf s ,t) = \xi_1$ and $\xi_2(\mathbf s ,t) = \xi_2$, we assume a Gamma prior distribution, which leads to the conditional posterior distribution for $\xi_1|\cdot$ and $\xi_2|\cdot$,
\begin{align*}
\mathbf \xi_1|\cdot & \propto \pi(\xi_1) \times \prod_{i,j}f(Y(s_i,t_j)\; ;W(s_i,t_j \;; \xi(s_i,t_j)),
\sigma^2) \tag{A3}\\
\mathbf \xi_2|\cdot & \propto \pi(\xi_2)\times\prod_{i,j}f(Y(s_i,t_j)\; ;W(s_i,t_j \;; \xi(s_i,t_j)),
\sigma^2)
\end{align*}
which are intractable.

2. For the full model where $\log\{\xi_1(\mathbf s,t)\}=\alpha_0 +\alpha_1 +I(\mathbf s,t)+\nu$ we assume (i) a point-mass prior distribution when $I(\mathbf s,t)=0$, $ \log\{\xi_1(\mathbf s,t)\}=\hat \xi_{1, I=0}$, (ii) when  $I(\mathbf s,t)=1$,  $ \log\{\xi_1(\mathbf s,t)\}= \log(\xi_{1, I=1})= \alpha_0 + \alpha_1 + \nu, \; \nu\sim \mathcal{N}(0, \sigma^2_{\nu}).$
\begin{enumerate}
    \item The conditional posterior distribution for $\xi_{1, I=1}$ is given by 
    \begin{align*}
        \mathbf \xi_{1, I=1}|\cdot & \propto f( \log(\xi_{1, I=1})\; ; \alpha_0 + \alpha_1, \sigma^2_{\alpha})\times\prod_{i,j}f(Y(s_i,t_j)\; ;W(s_i,t_j \;; \xi(s_i,t_j)) \tag{A4}
    \end{align*}
    \item Assuming a bivariate Gaussian prior distribution for $\mathbf \alpha$, 
    \begin{align*}
        \mathbf \alpha|\cdot \propto f(\mathbf \alpha \; ;\mathbf \mu_\alpha,\sigma^2_
        \alpha \mathbf I)\times \prod_{i,t}f( \log(\xi(s_i,t_j))\; ;\alpha_0 + \alpha_1, \sigma_\nu^2 ) \tag{A5}
    \end{align*}
    which leads to a conditionally Gaussian posterior distribution.
      \item Assuming an inverse-gamma prior distribution for $\sigma_\nu^2$,  
    \begin{align*}
        \sigma_\nu^2|\cdot \propto \mathcal  \pi(\sigma_\nu^2)\times \prod_{i,t}f( \log(\xi(s_i,t_j))\; ;\alpha_0 + \alpha_1, \sigma_\nu^2 ) \tag{A6}
    \end{align*}
    which leads to an inverse-gamma conditional posterior distribution.
    
\end{enumerate}

The MCMC algorithms are given in the tables below:

\begin{algorithm}
\SetAlgoLined
%\KwIn{Initial values $\mathbf \beta^{(0)},\xi_1^{(0)},\xi_2^{(0)},\sigma^{2(0)}$; iterations $T$; log-proposal variance $\delta_1$.}
%\KwOut{Samples $\{\mathbf \beta^{(t)},\xi_1^{(t)},\xi_2^{(t)},\sigma^{2(t)}\}_{t=1}^T$.}

\For{$t\leftarrow 1$ \KwTo $T$}{
  \textbf{1. Sample } $\mathbf\beta^{(t)}$ from $\;\mathbf\beta\mid\cdot$  (A1.1).\;
  \vspace{2mm}
\textbf{2. Sample } $\sigma_b^{2^{(t)}}$ from $\;\sigma_b^2\mid\cdot$  (A1.2).\;
  \vspace{2mm} 
  \textbf{3. For each } $m\in\{1,2\}$\; (random-walk Metropolis on log-scale):\;
  \begin{enumerate}
    \item Draw proposal on the log-scale
    \[
      \log(\xi_m') \sim \mathcal{N}\big(\log(\xi_m^{(t-1)}),\,\delta_1\big),
    \]
    and set $\xi_m'=\exp(\log(\xi_m'))$.
    \item Compute the Metropolis--Hastings acceptance probability using (A4):
    \[
      p \;=\; \min\!\left\{1,\;
        \frac{\pi(\xi_m' \mid \text{rest})}{\pi(\xi_m^{(t-1)} \mid \text{rest})}
        \cdot
        \frac{q(\xi_m^{(t-1)}\mid\xi_m')}{q(\xi_m'\mid\xi_m^{(t-1)})}
      \right\}.
    \]
    Because the proposal is Gaussian on the log scale,
    \[
      \frac{q(\xi_m^{(t-1)}\mid\xi_m')}{q(\xi_m'\mid\xi_m^{(t-1)})}
      \;=\;\frac{1/\xi_m^{(t-1)}}{1/\xi_m'} \;=\; \frac{\xi_m'}{\xi_m^{(t-1)}}.
    \]
    Hence the working form is
    \[
      p \;=\; \min\!\left\{1,\;
        \frac{\pi(\xi_m' \mid \text{rest})}{\pi(\xi_m^{(t-1)} \mid \text{rest})}
        \cdot \frac{\xi_m'}{\xi_m^{(t-1)}}
      \right\}.
    \]
    \item Draw $u\sim\mathcal{U}(0,1)$. If $u\le p$ then accept: $\xi_m^{(t)}\leftarrow\xi_m'$, else reject: $\xi_m^{(t)}\leftarrow\xi_m^{(t-1)}$, and update $W(\mathbf s, \tau ;\mathbf \xi^{(t)})\leftarrow W(\mathbf s, \tau ;\mathbf \xi^{(t-1)})$.
  \end{enumerate}
  \vspace{2mm}

  \textbf{4. Sample } $\sigma^{^2{(t)}}$ from $\;\sigma^{2}\mid\cdot$ (A2)
}
\caption{Metropolis-within-Gibbs when $\xi_m(\mathbf s,t)=\xi_m$}
\end{algorithm}

\begin{algorithm}
\SetAlgoLined
%\KwIn{Initial values $\mathbf \beta^{(0)},\xi_1^{(0)},\xi_2^{(0)},\sigma^{2(0)}$; iterations $T$; log-proposal variance $\delta_1$.}
%\KwOut{Samples $\{\mathbf \beta^{(t)},\xi_1^{(t)},\xi_2^{(t)},\sigma^{2(t)}\}_{t=1}^T$.}

\For{$t\leftarrow 1$ \KwTo $T$}{
  \textbf{1. Sample } $\mathbf\beta^{(t)}$ from $\;\mathbf\beta\mid\cdot$  (A1.1).\;
  \vspace{2mm}
\textbf{2. Sample } $\sigma_b^{2^{(t)}}$ from $\;\sigma_b^2\mid\cdot$  (A1.2).\;
  \vspace{2mm}
  
  \textbf{3.} Fix $\xi_2$ and $\xi_1(s,t)$ when $I(s,t)=0$. For $\xi_1(s,t)=\xi_{1, I=1}$ when $I(s,t)=1$ \; (random-walk Metropolis on log-scale):\;
  \begin{enumerate}
    \item Draw proposal on the log-scale
    \[
      \log(\xi_{1, I=1}') \sim \mathcal{N}\big(\log(\xi_{1, I=1}^{(t-1)}),\,\delta_1\big),
    \]
    and set $\xi_{1, I=1}'=\exp(\log(\xi_{1, I=1}'))$.
    \item Compute the Metropolis--Hastings acceptance probability using (A4):
    \[
      p \;=\; \min\!\left\{1,\;
        \frac{\pi(\xi_{1, I=1}' \mid \text{rest})}{\pi(\xi_{1, I=1}^{(t-1)} \mid \text{rest})}
        \cdot
        \frac{q(\xi_{1, I=1}^{(t-1)}\mid\xi_m')}{q(\xi_{1, I=1}'\mid\xi_{1, I=1}^{(t-1)})}
      \right\}.
    \]
    Because the proposal is Gaussian on the log scale,
    \[
      \frac{q(\xi_{1, I=1}^{(t-1)}\mid\xi_{1, I=1}')}{q(\xi_{1, I=1}'\mid\xi_{1, I=1}^{(t-1)})}
      \;=\;\frac{1/\xi_{1, I=1}^{(t-1)}}{1/\xi_{1, I=1}'} \;=\; \frac{\xi_{1, I=1}'}{\xi_{1, I=1}^{(t-1)}}.
    \]
    Hence the working form is
    \[
      p \;=\; \min\!\left\{1,\;
        \frac{\pi(\xi_{1, I=1}' \mid \text{rest})}{\pi(\xi_{1, I=1}^{(t-1)} \mid \text{rest})}
        \cdot \frac{\xi_{1, I=1}'}{\xi_{1, I=1}^{(t-1)}}
      \right\}.
    \]
    \item Draw $u\sim\mathcal{U}(0,1)$. If $u\le p$ then accept: $\xi_{1, I=1}^{(t)}\leftarrow\xi_{1, I=1}'$, else reject: $\xi_{1, I=1}^{(t)}\leftarrow\xi_{1, I=1}^{(t-1)}$, and update $W(\mathbf s, \tau ;\mathbf \xi^{(t)})\leftarrow W(\mathbf s, \tau ;\mathbf \xi^{(t-1)})$.
  \end{enumerate}
  \vspace{2mm}

  \textbf{4. Sample } $\sigma^{2^{(t)}}$ from $\;\sigma^2\mid\cdot$ (A2)
  
  \vspace{2mm}

  \textbf{5. Sample } $\mathbf \alpha^{{(t)}}$ from $\mathbf \alpha\mid\cdot$ (A5)
  
  \vspace{2mm}

  \textbf{6. Sample } $\mathbf \sigma_\nu^{2^{(t)}}$ from $\sigma_\nu^2\mid\cdot$ (A6)
}
\caption{Metropolis-within-Gibbs when $\log(\xi_1(\mathbf s,t))=\alpha_0 + \ \alpha_1I(s,t) +\nu$}
\end{algorithm}

\end{document}